\documentclass[12pt]{article}
\usepackage{a4}
\usepackage{amsmath}
\usepackage{amssymb}
\usepackage{amsfonts}
\usepackage{graphics}
\usepackage{epsfig}
\usepackage{latexsym}

\begin{document}

\noindent  arXiv:1210.7227
   \hfill  October 2012 \\

\renewcommand{\theequation}{\arabic{section}.\arabic{equation}}
\thispagestyle{empty}
\vspace*{-1,5cm}
\noindent \vskip3.3cm

\begin{center}
{\Large\bf Radial Reduction and Cubic Interaction for Higher Spins in $(A)dS$ space }
\vspace*{1 cm}

{\large Ruben Manvelyan ${}^{\dag\ddag}$, Ruben Mkrtchyan${}^{\ddag}$  and \\Werner R\"uhl${}^{\dag}$}

\medskip

${}^{\dag}${\small\it Department of Physics\\ Erwin Schr\"odinger Stra\ss e \\
Technical University of Kaiserslautern, Postfach 3049}\\
{\small\it 67653
Kaiserslautern, Germany}\\
\medskip
${}^{\ddag}${\small\it Yerevan Physics Institute\\ Alikhanian Br.
Str.
2, 0036 Yerevan, Armenia}\\
\medskip
{\small\tt manvel,ruehl@physik.uni-kl.de; mrl@web.am}
\end{center}\vspace{2cm}

\bigskip
\begin{center}
{\sc Abstract}
\end{center}
\quad
We present a new version of the radial reduction formalism to obtain a cubic interaction of higher spin gauge fields in $AdS_{d+1}$ space from the corresponding cubic interaction in a flat $d+2$ dimensional background. We modify the radial reduction procedure proposed previously by T. Biswas and W. Siegel in 2002 and applied to the free higher spin Lagrangian by K. Hallowell and A. Waldron in 2005. This modified radial reduction scheme is applied to interacting massless higher spin fields in Fronsdal's formulation, and all results are expressed in a direct $AdS_{d+1}$ invariant way with $AdS$ covariant derivatives. We present a consistent algorithm and define new procedure to obtain all corrections proportional to powers of the cosmological constant, and apply these to the main term of the cubic self-interaction.

\newpage

\section*{Introduction}
\quad
Although consistent equations of motion \cite{VasilievEqn} for interacting higher spin fields are known for many years, the action principle for these theories remains unknown. For recent reviews on the state of the art in higher spin gauge field theories see \cite{Vasiliev:2012vf}-\cite{Sagnotti:2011qp}. One can observe during the last years growing interest in this topic connected with recent progress in the construction of cubic interactions for higher spin gauge fields and its association to string theory \cite{Metsaev}-\cite{polyakov}, which gave new insights in earlier important works \cite{Bengtsson:1983pd}-\cite{Metsaev:1993ap}. In recent years cubic interactions of higher spin fields were discussed in the literature using different technics. For important recent works see \cite{Vasiliev2}-\cite{Alkalaev:2009vm} and references therein.

The fact that Vasiliev equations \cite{VasilievEqn} are naturally formulated on AdS background space-time, important developments of recent years connected with the Klebanov-Polyakov conjecture \cite{Kleb}\footnote{For the recent development see \cite{Giombi:2011ya}, \cite{Maldacena:2012sf}
and references therein.}, as well as recent progress in three dimensional $AdS$ higher spin gravity resulting in new relations between topological Chern-Simons theory, two dimensional conformal theories with higher spin symmetry, and new 3-dimensional black hole solutions with higher spin charges (\cite{Campoleoni:2011hg}-\cite{Kraus:2011ds} and references therein), point out the importance of an (A)dS background for the consistent (linearized) higher spin interacting theories.

Free Lagrangians for Higher Spin gauge fields both in flat space and in constantly curved backgrounds being known over thirty years \cite{Frons}, attempts to construct Lagrangians for interacting theories haven't been successful yet beyond the cubic vertices.
For the symmetric higher spin fields all the covariant cubic vertices in Minkowski background are available now \cite{MMR1}. This vertices were derived and classified in light cone gauge by Metsaev \cite{Metsaev}.

  In \cite{MMR1} all cubic interactions for higher spin fields were derived in a covariant off-shell formulation in full agreement with light cone gauge results of Metsaev \cite{Metsaev} and show that all interactions of higher spin gauge fields with any spins $s_{1}, s_{2}, s_{3}$ both in flat space and in dS or AdS are unique up to partial integration and field redefinition\footnote{This was already proven for some low spin cases of both the Fradkin-Vasiliev vertex for $2, s, s$ and the nonabelian vertex for $1, s, s$  in \cite{boulanger}.} therefore vertices of \cite{MMR1} can reproduce the flat limit of well-known Fradkin-Vasiliev AdS vertices \cite{Vasiliev}.

Now we turn to an opposite task, such as the derivation of higher spin interactions in $AdS$ space from known answers in a flat space of one dimension higher. We should mention here two different approaches : The first one was developed for the free case in the well known papers of Fronsdal \cite{Frons} and grounds on the so-called "ambient" space approach. The main idea and the methods of this approach is the realization of the $AdS_{D}$ space as a hypersphere in a corresponding $D+1$ dimensional flat space. Corresponding on-shell considerations in the interacting case were performed in \cite{Joung:2011ww}, \cite{Joung:2012rv} for both (partially) massless and massive cases. This consideration allows to apply the ansatz and Noether's procedure in the $AdS$ background in a similar way as in the flat space case, but on the other hand does not lead in a straightforward fashion to expressing the answers in explicitly gauge invariant and $AdS$ covariant ways with $AdS$ covariant derivatives.

The second approach is  connected with a radial reduction procedure proposed
ten years ago by T. Biswas and W. Siegel \cite{BS}. This approach includes a Kaluza-Klein expansion and a special procedure of separation of "massive" modes in order to connect free actions in flat $d+2$ dimensional space and constant curvature $d+1$ dimensional (A)dS background. This approach was developed for the free higher spin case in \cite{Waldron} by K. Hallowell and A. Waldron and leads to some progress in formulating free partially massless higher spin theories.

In this article we consider the latter approach for linearized massless higher spin theories including a local cubic interaction and develop some modified radial dimensional reduction scheme avoiding the problem with the Kaluza-Klein expansion that is not so effective in the interacting case and produces uncontrollable interactions between different (Stueckelberg) fields. In other words the existence of additional fields produces difficulties to go from one interacting Fronsdal double traceless tensor field in flat space to one double traceless higher spin gauge field with the same spin in an $AdS$ space of one dimension less. The main goal of this modification is to reproduce the standard $AdS_{d+1}$ gauge invariance for the spin $s$ field of the higher spin theory from the field of the same spin in $d+2$ dimensional flat background. This allows us to formulate an algorithm and corresponding recursion relations for getting all $AdS_{d+1}$ corrections to the cubic interaction in explicit $AdS$ covariant and gauge invariant manner.
This approach proves is applicable for the minimal selfinteraction.

In the first section we formulate the modified radial dimensional reduction scheme and apply it to the free case comparing with the corresponding considerations in \cite{BS} and \cite{Waldron}.

In the second section we consider the main term of the cubic interaction developed in \cite{MMR1} and \cite{MMR2}. We rewrite this term in the form convenient for analyzing it under the aspect of radial reduction. We show that scaling properties of the cubic (self)interaction in the case of a minimal number of derivatives are relevant for performing the radial dimensional reduction in a noncontradictory way.

In section three we check the proposed radial reduction for the already known case of spin two in all details. In section four we apply the reduction and obtain a first correction term to the main part of the cubic self-interaction for the general spin $s$ case using information obtained from the spin two case. Finally in section five we succeeded to derive all curvature correction terms for the main term of the cubic self-interaction for a spin $s$ field. In addition we present an appendix where we develop a technic for an expansion of powers of flat space symmetrized derivatives in curvilinear coordinates in terms of power series of covariant derivatives in an $AdS$ space of one dimension less. This technic can be useful in the future for obtaining all corrections proportional to powers of the cosmological constant or inverse  $AdS_{d+1}$ radius for all other terms of the interaction proportional to divergences and traces of higher spin gauge fields which are left for a separate publication.

\section{Radial reduction in the massless case and free HS gauge fields in $(A)dS$ }\setcounter{equation}{0}
\quad We start from a short review of the radial reduction technique developed in \cite{BS} and applied in detail to the free higher spin case in \cite{Waldron}. First we should introduce the following conventions.
As usual we utilize instead of symmetric tensors such as $h^{(s)}_{\mu_1\mu_2...\mu_s}(z)$ polynomials homogeneous in the vector $a^{\mu}$ of degree $s$ at the base point $z$
\begin{equation}
h^{(s)}(z;a) = \sum_{\mu_{i}}(\prod_{i=1}^{s} a^{\mu_{i}})h^{(s)}_{\mu_1\mu_2...\mu_s}(z) .\label{1.1}
\end{equation}
Then we can write the symmetrized gradient, trace, and divergence \footnote{To distinguish easily between "$a$" and "$z$" spaces we introduce the notation $\nabla_{\mu}$ for space-time derivatives $\frac{\partial}{\partial z^{\mu}}$.}
\begin{eqnarray}
&&Grad:h^{(s)}(z;a)\Rightarrow Gradh^{(s+1)}(z;a) = (a\nabla)h^{(s)}(z;a) , \label{1.2}\\
&&Tr:h^{(s)}(z;a)\Rightarrow Trh^{(s-2)}(z;a) = \frac{1}{s(s-1)}\Box_{a}h^{(s)}(z;a) ,\label{1.3}\\
&&Div:h^{(s)}(z;a)\Rightarrow Divh^{(s-1)}(z;a) = \frac{1}{s}(\nabla\partial_{a})h^{(s)}(z;a) .\label{1.4}
\end{eqnarray}
Moreover we introduce the notation $*_a, *_b,\dots$ for a contraction in the symmetric spaces of indices $a$ or $b$
\begin{eqnarray}
  *_{a}&=&\frac{1}{(s!)^{2}} \prod^{s}_{i=1}\overleftarrow{\partial}_{a^{\mu_{i}}}\overrightarrow{\partial}_{a_{\mu_{i}}} .\label{1.5}
\end{eqnarray}
In this article we distinguish between flat space coordinates/derivatives and $AdS$ space coordinates/derivatives as  $X^{A}$ /$\tilde{\nabla}$ and $x^{\mu}$/$\nabla$ correspondingly, keeping ordinary $\partial_{a^{M}}$ and $\partial_{a^{\mu}}$ for operations in the auxiliary vector space.
The main goal of radial reduction described in \cite{BS} and \cite{Waldron} is the reconsideration of the free field theory in the flat $d+2$ dimensional space
\begin{eqnarray}
&&X^{A} \quad A=1,2,.\dots d+2 ,\label{1.6}\\
&& ds^{2}=\eta_{AB}dX^{A}dX^{B}=-(dX^{d+2})^{2}+(dX^{d+1})^{2}+dX^{i}dX^{j}\eta_{ij} , \label{1.7}
\end{eqnarray}
in a curvilinear coordinate system defined by the following coordinate transformations $X^{A}\rightarrow (u,r,x^{i})$:
\begin{eqnarray}
  X^{d+2} &=& \frac{1}{2} e^{u}[r+\frac{1}{r}(\pm L^{2}+x^{i}x^{j}\eta_{ij})] , \label{1.8}\\
  X^{d+1}&=& \frac{1}{2}e^{u}[r-\frac{1}{r}(\pm L^{2}-x^{i}x^{j}\eta_{ij})] , \label{1.9}\\
  X^{i} &=& e^{u}L\frac{x^{i}}{r}, \label{1.10}\\
  \mp e^{2u} L^{2} &=& -(X^{d+2})^{2}+(X^{d+1})^{2}+ X^{i}X^{j}\eta_{ij} ,\label{1.11}\\
  ds^{2}&=& L^{2}e^{2u}[\mp du^{2}+\frac{1}{r^{2}}(\pm dr^{2} + dx^{i}dx^{j}\eta_{ij})] .\label{1.12}
\end{eqnarray}
It is easy to see that after restricting $e^{u}=1$ we obtain instead of coordinate transformations the usual embedding of the $(A)dS_{d+1}$ hypersphere into $d+2$ dimensional flat space ($AdS$ for upper signs in $\pm$ and $\mp$). From now on we work in the Euclidian version of $AdS_{d+1}$ space for simplicity with unit radius, which means we choose $\eta_{ij}=\delta_{ij}$ and $L=1$.

In other words we should formulate $d+2$ dimensional field theory in the curvilinear coordinates with flat
$e^{2u}(AdS_{d+1}\times \mathcal{R}_{u})$ metric
\begin{equation}\label{1.13}
    ds^{2}=e^{2u}[-du^{2}+ g_{\mu\nu}(x) dx^{\mu}dx^{\nu}]=G_{uu}(u)du^{2} +G_{\mu\nu}(u,x)dx^{\mu}dx^{\nu} ,
\end{equation}
where $x^{\mu}=(r, x^{i})$ are $AdS_{d+1}$ coordinates and $g_{\mu\nu}(x)=\frac{1}{r^{2}}\delta_{\mu\nu}$ is the corresponding constant curvature metric. In this curvilinear coordinate we use corresponding transformed tensors and commuting (because space is still flat) covariant derivatives $\tilde{\nabla}_{A}$ constructed for the metric (\ref{1.13}) with flat connection (Christoffel symbols) $\Gamma^{C}_{AB}$ \begin{eqnarray}
  \Gamma^{u}_{uu} &=& 1 , \quad \Gamma^{u}_{\mu\nu} = g_{\mu\nu} , \quad \Gamma^{\mu}_{u\nu}=\delta^{\mu}_{\nu}, \quad\Gamma^{u}_{u\mu}=\Gamma^{\mu}_{uu}=0 ,\label{Christ1}\\
\Gamma^{\mu}_{\nu\lambda}
  &=&\Gamma^{\mu(AdS)}_{\nu\lambda}=\frac{1}{2}g^{\mu\rho}(\partial_{\nu}g_{\lambda\rho}
  +\partial_{\lambda}g_{\nu\rho}-\partial_{\rho}g_{\nu\lambda}) , \label{Christ2}
  \end{eqnarray}
These covariant derivatives work in $X^{A}=(u, x^{\mu})$ space and are not connected primarily with our auxiliary vectors $a^{A}$ inserted only for shortening symmetric tensor contractions and symmetrizing procedures. For instance the action of the covariant derivatives in curvilinear space is just like in Cartesian case:
\begin{equation}
    \tilde{\nabla}_{A}h^{(s)}(X;a)=(\tilde{\nabla}_{A}h_{A_{1}A_{2}\dots A_{s}})a^{A_{1}}a^{A_{2}}\dots a^{A_{s}}. \label{cond}
\end{equation}
But now our covariant derivatives involves all Christoffel symbols (\ref{Christ1}), (\ref{Christ2}). Using the exact form of them we see that we can realize our covariant derivatives working on rank $s$ symmetric tensors as operators working in both $X$ and $a$ spaces in the following way
\begin{eqnarray}
    {\tilde{\nabla}}_{A} &=& (\nabla_{u}, D_{\mu}) ,\label{1.14}\\
     \nabla_{u}&=& \partial_{u}-a^{u}\partial_{a^{u}}-a^{\mu}\partial_{a^{\mu}} ,\label{1.15}\\
    D_{\mu} &=& \nabla_{\mu}-a^{u}\partial_{a^{\mu}}-a_{\mu}\partial_{a^{u}} ,\label{1.16}
  \end{eqnarray}
where $\nabla_{\mu}=\nabla^{AdS}_{\mu}$ is the $AdS_{d+1}$ covariant derivative defined by Christoffel symbols (\ref{Christ2}) and all other Christoffel symbols (\ref{Christ1}) are realized through the differential operators in auxiliary $a$ space. After this representation of the covariant derivatives the condition (\ref{cond}) will hold only for ordinary derivative $\partial_{u}$ and $AdS_{d+1}$ derivatives $\nabla_{\mu}$
\begin{eqnarray}
\partial_{u}h^{(s)}(X;a)&=&(\partial_{u}h_{A_{1}A_{2}\dots A_{s}})a^{A_{1}}a^{A_{2}}\dots a^{A_{s}} ,\\
  \nabla_{\mu}h^{(s)}(X;a)&=&(\nabla_{\mu}h_{A_{1}A_{2}\dots A_{s}})a^{A_{1}}a^{A_{2}}\dots a^{A_{s}} .
\end{eqnarray}
This $AdS_{d+1}$ covariant representation for derivatives gives us a simple tool for radial reduction of tensors and derivatives and extraction of $AdS$ corrections in an elementary algebraic way.

Our final prescription for radial reduction in the massless $AdS$ case slightly differs from the reduction formulated in   \cite{BS} and \cite{Waldron} for general massive and (partially) massless cases and can be summarized by the following three points.
\begin{enumerate}
  \item Rewrite all derivatives in the curvilinear  coordinates with metric
  (\ref{1.13}) in the way (\ref{1.14})-(\ref{1.16})
  \item Restrict all double traceless higher spin gauge fields to corresponding double traceless tensors in $AdS$ space of dimension one less
      \begin{eqnarray}
       h^{(s)}(X^{A};a^{A})&\equiv& a^{A_{1}}a^{A_{2}}\dots a^{A_{s}} h^{(s)}_{A_{1}A_{2}\dots A_{s}}(u,x^{\mu}) =>  h^{(s)}(u,x^{\mu};a^{\mu}) ,\quad\quad \label{1.17}\\
       h^{(s)}(u,x^{\mu};a^{\mu}) &\equiv& h^{(s)}(X^{A};a^{A})|_{a^{u}=0}= a^{\mu_{1}}a^{\mu_{2}}\dots a^{\mu_{s}}h^{(s)}_{\mu_{1}\mu_{2}\dots \mu_{s}}(u,x^{\mu}) .\quad\quad \label{1.18}
      \end{eqnarray}
      \item Restrict the dependence on additional "$u$" coordinates for all fields and gauge parameters in the following way
       \begin{eqnarray}
        h^{(s)}(u,x^{\mu};a^{\mu}) &=& e^{2(s-1)u}h^{(s)}(x^{\mu};a^{\mu}) .\label{1.19}
      \end{eqnarray}
      \end{enumerate}

 Actually the last two points replace two steps performed in \cite{BS} and \cite{Waldron} where those authors formulated the following:
 \begin{itemize}
      \item Instead of (\ref{1.14}) these authors solve the double traceless condition for a $d+2$ spin $s$ field and obtain instead a set of four unconstrained $d+1$ dimensional tensor fields with spins $s, s-1, s-2, s-3$. Then they perform a Weyl rescaling of fields
      \begin{eqnarray}
        h^{(s)}(u,x^{\mu};a^{\mu}) &=>& e^{(s-d/2)u}h^{(s)}(u,x^{\mu};a^{\mu}) , \label{1.20}
      \end{eqnarray}
       to split $AdS$ and $u$  dependence in the free action.
      \item After that in \cite{Waldron} the authors make a Wick rotation of the $u$ coordinates
      \begin{eqnarray}
       u &=>& i u ,\label{1.21}\\
       \partial_{u} &=>&-i \partial_{u} ,\label{1.22}
       \end{eqnarray}
      and after compactification of the $u$ coordinate they perform a Fourier expansion of the gauge field
      \begin{equation}
        h^{(s)}(u,x^{\mu};a^{\mu})=\sum_{m=-\infty}^{+\infty}e^{i mu}h^{(s)}_{m}(x^{\mu};a^{\mu}) .\label{1.23}
      \end{equation}
      In order to obtain only one gauge field in $AdS_{d+1}$ for one massless spin $s$, one should perform a truncation to the              separate special mode
      \begin{equation}
        m=s+d/2-2 .\label{1.24}
      \end{equation}
\end{itemize}

Now we illustrate and comment on our procedure (\ref{1.17})-(\ref{1.19}) considering the case of a free higher spin gauge field theory and on the difference between this approach and the approach developed in \cite{Waldron} for the general case including (partial) massive theories.

First of all we use Fronsdal's formulation, and our initial field in $d+2$ dimensional flat space is double traceless
\begin{eqnarray}
  \Box_{a^{A}}^{2}h^{(s)}(X^{A};a^{A})&=& 0 .\label{1.25}
\end{eqnarray}
Rewriting this condition in the curvilinear coordinates (\ref{1.13})
\begin{eqnarray}
  e^{-4u}[-\partial^{2}_{a^{u}}+g^{\mu\nu}(x)\partial_{a^{\mu}}\partial_{a^{\nu}}]^{2}h^{(s)}(u,x^{\mu};a^{u},a^{\mu}) &=& 0 ,\label{1.26}
\end{eqnarray}
we see that our first condition for radial reduction of fields (\ref{1.17})-(\ref{1.18})
leads to the correct double tracelessness condition in $d+1$ dimensional $AdS$ space with metric $g^{AdS}_{\mu\nu}(x)=g_{\mu\nu}(x)$
\begin{equation}\label{1.27}
    \Box_{a^{\mu}}^{2}h^{(s)}(u,x^{\mu};a^{\mu})
    =[g^{\mu\nu}(x)\partial_{a^{\mu}}\partial_{a^{\nu}}]^{2}h^{(s)}(u,x^{\mu};a^{\mu})=0 ,
\end{equation}
for the one parameter (namely $u$ dependent) family of $AdS_{d+1}$ symmetric double traceless gauge fields $h^{(s)}(u,x^{\mu};a^{\mu})$. So we see that maintaining the condition of double tracelessness from the $d+2$ dimensional case to the $AdS_{d+1}$ case, we deal with \emph{only one Fronsdal field} and get rid of any additional Stueckelberg degree of freedom needed for the gauge invariant formulation in the massive case (see \cite{Waldron}) that is completely useless in the massless gauge field case considered in this article.

Now we turn to the scaling behavior of the free higher spin Fronsdal action
\begin{eqnarray}
  S_{0}[h^{(s)}(X^{A};a^{A})]&=& \int d^{d+2}X \mathcal{L}_{0}(h^{(s)}(X^{A};a^{A})) \label{1.28}
\end{eqnarray}
where Fronsdal's Lagrangian is:
\begin{eqnarray}
 \mathcal{L}_{0}(h^{(s)}(X^{A};a^{A}))&=&-\frac{1}{2}h^{(s)}(X^{A};a^{A})*_{a^{A}}\mathcal{F}^{(s)}(X^{A};a^{A})
    \nonumber\\&+&\frac{1}{8s(s-1)}\Box_{a^{A}}h^{(s)}(X^{A};a^{A})*_{a^{A}}\Box_{a^{A}}\mathcal{F}^{(s)}(X^{A};a^{A}) ,\label{1.29}
\end{eqnarray}
$\mathcal{F}^{(s)}(X^{A};a^{A})$ is the Fronsdal tensor
\begin{eqnarray}
&&\mathcal{F}^{(s)}(X^{A};a^{A})=\Box_{d+2} h^{(s)}(X^{A};a^{A})-s(a^{A}\tilde{\nabla}_{A})D^{(s-1)}(X^{A};a^{A}) , \quad\label{1.30}
\end{eqnarray}
and $D^{(s-1)}(X^{A};a^{A})$ is the deDonder tensor or traceless divergence of the higher spin gauge field
\begin{eqnarray}
 &&s D^{(s-1)}(X^{A};a^{A}) = (\tilde{\nabla}^{A}\partial_{a^{A}})h^{(s)}(X^{A};a^{A})
-\frac{1}{2}(a^{A}\tilde{\nabla}_{A})\Box_{a^{A}}h^{(s)}(X^{A};a^{A}) ,\quad\label{1.31}\\
&& \Box_{a^{A}} D^{(s-1)}(X^{A};a^{A})=0 .\nonumber
\end{eqnarray}
The initial gauge variation of order zero in the spin $s$ field is
\begin{eqnarray}\label{1.32}
\delta_{(0)} h^{(s)}(X^{A};a^{A})=s (a^{A}\tilde{\nabla}_{A})\epsilon^{(s-1)}(X^{A};a^{A}) ,
\end{eqnarray}
with the traceless gauge parameter for the double traceless gauge field
\begin{eqnarray}
&&\Box_{a^{A}}\epsilon^{(s-1)}(X^{A};a^{A})=0 ,\label{1.33}
\end{eqnarray}
From (\ref{1.28})-(\ref{1.31}) we see that scaling behaviour of all terms in the free action is the same as in the first term with  the Laplacian, so we need only analyze the main term
\begin{equation}\label{1.34}
 \int d^{d+2}X  \Big(h^{(s)}_{A_{1}A_{2}\dots A_{s}}\tilde{\nabla}_{A}\tilde{\nabla}_{B}h^{(s)}_{B_{1}B_{2}\dots B_{s}}\eta^{AB}\eta^{A_{1}B_{1}}\dots\eta^{A_{s}B_{s}} + \dots\Big) .
\end{equation}
In curvilinear coordinates with metric (\ref{1.13}) and after restriction of the fields (\ref{1.17}) we obtain
\begin{eqnarray}
 &&\int du e^{(d-2s)u} d^{d+1}x \sqrt{g(x)}\nonumber \\
  &&\times\Big(h^{(s)}_{\mu_{1}\mu_{2}\dots \mu_{s}}(u,x)(\nabla^{2}_{u}+g^{\mu\nu}(x)\nabla_{\mu}\nabla_{\nu})h^{(s)}_{\nu_{1}\nu_{2}\dots \nu_{s}}(u,x)g^{\mu_{1}\nu_{1}}(x)\dots g^{\mu_{s}\nu_{s}}(x) + \dots\Big),\quad\quad\quad\label{1.35}
\end{eqnarray}
and we see that the Weyl rescaling (\ref{1.20}) of \cite{Waldron} removes the factor $e^{(d-2s)u}$ in the measure and prepares the free action for the Kaluza-Klein procedure (\ref{1.21}), (\ref{1.22}). Instead of this Weyl rescaling and hereupon the following Kaluza-Klein mode extraction we propose the general $u$ dependence restriction (\ref{1.19}), more suitable in this massless case.

To understand that, we focus now on the free gauge transformation (\ref{1.32}). Next we apply  the same type of reduction condition to the gauge parameter
\begin{eqnarray}
  \epsilon^{(s-1)}(X^{A};a^{A}) &=>& \epsilon^{(s-1)}(u,x;a^{u},a^{\mu})|_{a^{u}=0} =\epsilon^{(s-1)}(u,x;a^{\mu}),\label{1.36}
\end{eqnarray}
which maps the $d+2$ dimensional tracelessness condition (\ref{1.33}) onto the $d+1$ dimensional one
\begin{equation}\label{1.37}
    g^{\mu\nu}(x)\partial_{a^{\mu}}\partial_{a^{\nu}}\epsilon^{(s-1)}(u,x;a^{\mu})=\Box_{a^{\mu}}\epsilon^{(s-1)}(u,x;a^{\mu})=0 .
\end{equation}

Then we see that the Weyl rescaling of gauge fields (\ref{1.20}) and the transformation rule (\ref{1.32}) require
the same type of Weyl rescaling for the gauge parameter
\begin{equation}\label{1.38}
   \epsilon^{(s-1)}(u,x;a^{\mu})=> e^{(s-d/2)u}\epsilon^{(s-1)}(u,x;a^{\mu}) .
\end{equation}
So finally applying (\ref{1.14})-(\ref{1.19}), (\ref{1.36}) and (\ref{1.38}) to (\ref{1.32}), we obtain the following expression
\begin{eqnarray}
 &&\delta_{(0)} h^{(s)}(u,x;a^{\mu})=s e^{(d/2-s)u} (a^{u}\nabla_{u}+a^{\mu}D_{\mu})e^{(s-d/2)u}\epsilon^{(s-1)}(u,x;a^{\mu})\nonumber\\
 &&=s[ a^{\mu}\nabla_{\mu}+a^{u}(\partial_{u}-s-d/2+2)]\epsilon^{(s-1)}(u,x;a^{\mu}) .\label{1.39}
\end{eqnarray}
Analyzing the last expression we see that the second term in the bracket vanishes when we choose the "massless" mode (\ref{1.24}). So we see that for the special mode $m=s+d/2-2$ of the gauge field $ h^{(s)}(x;a^{\mu})=h^{(s)}_{(m)}(x;a^{\mu})$ and gauge parameter $\epsilon^{(s-1)}(x;a^{\mu})=\epsilon^{(s-1)}_{(m)}(x;a^{\mu})$, radial reduction restores the usual gauge transformation for a free $AdS_{d+1}$ massless gauge field from \ref{1.39}
\begin{equation}\label{1.40}
    \delta_{(0)} h^{(s)}_{(s+d/2-2)}(x;a^{\mu})=s(a^{\mu}\nabla_{\mu})\epsilon^{(s-1)}_{(s+d/2-2)}(x;a^{\mu}) ,
\end{equation}
which we cannot say about the gauge transformation of the complex conjugate mode $h^{(s)}_{(-m)}(x;a^{\mu})$ with gauge parameter $\epsilon^{(s-1)}_{(-m)}(x;a^{\mu})$. On the other hand the reality condition for the field $h^{(s)}(x;a^{\mu})$ leads to the standard  condition $h^{(s)*}_{m}=h^{(s)}_{-m}$ for the Kaluza-Klein expansion (\ref{1.23}), and therefore the natural restriction $h^{(s)*}_{m}=h^{(s)}_{m}$ to only one mode in the reduced action leads to a contradiction with the gauge transformation rule. In other words we cannot satisfy the first order differential equation $(\partial_{u}-s-d/2+2)\epsilon^{(s-1)}(u,x;a^{\mu})=0$ expanding the real parameter $\epsilon^{(s-1)}(u,x;a^{\mu})$
in $sin(mu)$ and $cos(mu)$.

This would force us to switch on another component of the initial $d+2$ dimensional tensor
$h^{(s)}(u,x;a^{u},a^{\mu})=\phi^{(s-1)}_{u}(u,x;a^{\mu})a^{u}=\phi^{(s-1)}(u,x)_{u\mu_{1}\dots\mu_{s-1}}a^{u}a^{m_{1}}\dots a^{\mu_{s-1}}$ to compensate the second term in the second line of (\ref{1.39}) as it was done in \cite{Waldron}. But in that case we arrive at the set of four unconstrained fields and two unconstrained parameters in $AdS_{d+1}$ instead of one double traceless gauge field and one traceless parameter. To get correctly the action for one massless field on $AdS_{d+1}$, one should after the reduction to four unconstrained fields obtain a consistent truncation to one double traceless field, which is more or less clear in the quadratic free case but
remains completely obscure in the case of an interaction. To overcome this difficulty we propose instead of Weyl rescaling and the Kaluza-Klein mode extraction, just a simple restriction of the "$u$" dependence for the gauge field (\ref{1.19}) and the same for the gauge parameter:
\begin{eqnarray}
    \epsilon^{(s-1)}(u,x;a^{\mu})= e^{2(s-1)u}\epsilon^{(s-1)}(x^{\mu};a^{\mu}) .\label{1.41}
\end{eqnarray}
Note that the exponential factor in the above formula and in (\ref{1.19}) is the effective value of the partial derivative $\partial_{u}$ including Weyl rescaling (\ref{1.20}), shifting all $u$ derivatives by $s-d/2$, and the exponential factor of the corresponding Kaluza-Klein mode (\ref{1.24}) which together with the Wick rotation (\ref{1.21}), (\ref{1.22}) shifts $u$ derivatives by $s+d/2-2$. So we see that on the level of the free equation of motion we get the same answer as for the Weyl rescaled Kaluza-Klein mode, but we avoid the discrepancy in the gauge transformation role (\ref{1.39}) which is without Weyl rescaling, and the Kaluza-Klein expansion looks now as:
\begin{eqnarray}\label{1.42}
 \delta_{(0)} h^{(s)}(u,x;a^{\mu})&=&s[ a^{\mu}\nabla_{\mu}+a^{u}(\partial_{u}-2(s-1))]\epsilon^{(s-1)}(u,x;a^{\mu}) .
\end{eqnarray}
Therefore we see that the restrictions (\ref{1.19}) and (\ref{1.41}) remove any $u$ dependence in (\ref{1.42}) and restore the correct gauge transformation for the remaining massless double traceless Fronsdal gauge field $h^{(s)}(x;a^{\mu})$
\begin{eqnarray}\label{1.43}
 \delta_{(0)} h^{(s)}(x;a^{\mu})&=&s a^{\mu}\nabla_{\mu}\epsilon^{(s-1)}(x;a^{\mu})
\end{eqnarray}
without problems with reality and with transformations of additional components.

In the same fashion one can prove that the gauge invariant Fronsdal tensor (\ref{1.30}) reduces after restriction (\ref{1.19}) to the ordinary Fronsdal tensor in the $AdS_{d+1}$ background
\begin{eqnarray}
 \mathcal{F}^{(s)}(X^{A};a^{A})&=& e^{2(s-2)u}\mathcal{F}^{(s)}(x;a^{\mu}) ,\label{1.44}
\end{eqnarray}
where
\begin{eqnarray}
  &&\mathcal{F}^{(s)}(x;a^{\mu})=\Box_{d+1} h^{(s)}(x^{\mu};a^{\mu})\qquad \nonumber\\
 && -(a^{\mu}\nabla_{\mu})\Big[(\nabla^{\nu}\partial_{a^{\nu}})h^{(s)}(x;a^{\mu})
-\frac{1}{2}(a^{\nu}\nabla_{\nu})\Box_{a^{\mu}}h^{(s)}(x;a^{\mu})\Big]\quad\quad\nonumber\\
&& -[s^{2}+s(d-5)-2(d-2)]h^{(s)}(x^{\mu};a^{\mu}))-g_{\mu\nu}a^{\mu}a^{\nu}h^{(s)}(x^{\mu};a^{\mu}).\label{1.45}
\end{eqnarray}
This tensor is gauge invariant in respect to (\ref{1.43}) and its last two terms are proper $1/L^2$ corrections in the $AdS_{d+1}$ background.
Moreover the "Bianchi" identity for the Fronsdal tensor being of the same form as in the flat case
\begin{eqnarray}
  (\nabla^{\nu}\partial_{a^{\nu}})\mathcal{F}^{(s)}(x;a^{\mu})
-\frac{1}{2}(a^{\nu}\nabla_{\nu})\Box_{a^{\mu}}\mathcal{F}^{(s)}(x;a^{\mu})=0 ,\label{1.46}
\end{eqnarray}
immediately proves that our $d+2$ dimensional flat space free action described by (\ref{1.28})-(\ref{1.31}) reduces after our radial reduction presented above to the proper free Fronsdal action in the $d+1$ dimensional constant curvature background
\begin{eqnarray}
  S_{0}[h^{(s)}(X^{A};a^{A})]&=&\left[\int du e^{(d+2s-4)u}\right]\times S[h^{(s)}(x^{\mu};a^{\mu})] ,\label{1.47}
\end{eqnarray}
where
\begin{eqnarray}
 S[h^{(s)}(x^{\mu};a^{\mu})]&=&\int d^{d+1}x\sqrt{g}\Big[-\frac{1}{2}h^{(s)}(x;a^{\mu})*_{a^{\mu}}\mathcal{F}^{(s)}(x;a^{\mu})
    \nonumber\\&+&\frac{1}{8s(s-1)}\Box_{a^{\mu}}h^{(s)}(x;a^{\mu})*_{a^{\mu}}\Box_{a^{\mu}}\mathcal{F}^{(s)}(x;a^{\mu})\Big] ,\label{1.48}
\end{eqnarray}
The overall infinite factor
\begin{equation}\label{1.49}
    \left[\int du e^{(d+2s-4)u}\right] ,
\end{equation}
will be discussed in the subsequent section.

\section{Cubic interactions of spin 2 fields and Radial Reduction to $AdS_{d+1}$}
\setcounter{equation}{0}\label{2.1}
In this section we consider the spin $2$ case example and check the radial reduction proposal for the linearized Einstein-Hilbert action up to the cubic level of interaction.
First of all we expand the Einstein-Hilbert action without cosmological term
\begin{equation}
    S^{d+2}=\int d^{d+2}\sqrt{G}R(G) ,
\end{equation}
around the flat background
\begin{equation}
G_{MN}=\eta_{MN}+h_{_{MN}} ,\label{2.2}
\end{equation}
and obtain a corresponding linearized expansion for the Lagrangian
\begin{equation}\label{2.3}
    S^{d+2}=\int(L_2 (h)+L_{3}(h)+O(h^{4})) .
\end{equation}
Here\footnote{We use in this section the notation $h=h^{M}_{M}$ for the trace of the spin 2 field and $(\partial h)^{M}$ for the divergence in a flat background, and correspondingly $h=h^{\mu}_{\mu}$ and $(\nabla h)^{\mu}$ for an $AdS$ background}
\begin{eqnarray}
{L_2(h)} =-\frac{1}{2}( \frac{1}{2}{(\partial_{R}h_{MN})(\partial^{R}h^{MN})} - {(\partial h)_M }{(\partial h)^M } + {(\partial h)_M }{\partial ^M }h - \frac{1}{2}({\partial _M }h)({\partial ^M }h)) ,\label{2.4}\quad\quad
\end{eqnarray}
and
\begin{eqnarray}
&& {{L}_{3}(h)}=-\frac{1}{2}(h[{{(\partial h)}_{M}}{{(\partial h)}^{M}}-\frac{3}{2}{\partial_{R}{h}_{M N }}{\partial^{N}{h}^{M R }}+\frac{1}{4}{\partial_{R}{h}_{M N}}{\partial^{R}{h}^{M N }} \nonumber\\  &&+\frac{1}{2}{{(\partial h)}^{M }}{\partial_{M}{h}}-\frac{1}{4}{\partial_{M}{h}}{\partial^{M}{h}}]
-2{\partial_{M}{h}}{\partial^{S}{h}^{M R}}{{h}_{R S }}+2h_{N }^{M }{\partial^{S}{h}^{N R}}\partial_{M}{{h}_{R S}} \nonumber\\&&+h_{N}^{M}{\partial_{R}{h}_{M S}}{\partial^{S}{h}^{N R}}-h_{N}^{M}{\partial_{S}{h}_{M R }}{\partial^{S}{h}^{N R}}-\frac{1}{2}h_{N}^{M}{\partial_{M}{h}_{R S}}{\partial^{N}{h}^{R S}}\nonumber\\
 && +\frac{1}{2}h_{N}^{M }{\partial_{M}{h}}{\partial^{N}{h}}-{{(\partial h)}^{M}}{\partial_{M}{h}_{R S}}{{h}^{R S}}+{\partial^{M}{h}}{\partial_{M}{h}_{R S}}{{h}^{R S}}) ,\label{2.5}
       \end{eqnarray}
where the derivatives were ordered in a way to get not more than one derivative acting on one spin two field.

Then we turn to $d+1$ dimensional gravitational action with cosmological constant\footnote{In our notation $S^{d+1}=\int d^{d+1}\sqrt{G}[R(G)-\Lambda]$ where $\Lambda=\frac{d(d-1)}{L^{2}}$, but from now on we put the radius $L=1$}
\begin{equation}
 S^{d+1}=\int d^{d+1}\sqrt{G}[R(G)-d(d-1)] , \label{2.6}
\end{equation}
and expand around a fixed $AdS_{d+1}$ background metric $g_{\mu\nu}=\frac{1}{r^{2}}\delta _{\mu \nu }$
\begin{equation}
G_{\mu\nu}=g_{\mu\nu}+h_{_{\mu\nu}} .\label{2.7}
\end{equation}
The resulting expansion
\begin{equation}\label{2.8}
    S^{d+1}=\int(L^{AdS}_2 (h)+L^{AdS}_{3}(h)+ O(h^{4})) ,
\end{equation}
produces quadratic
\begin{eqnarray}
&&L^{AdS}_2 (h) =
 - \frac{1}{2}\sqrt { - g} [\frac{1}{2}({\nabla _\rho }{h_{\mu \nu }})({\nabla ^\rho }{h^{\mu \nu }}) - ({\nabla ^\sigma }{h_{\mu \sigma }})({\nabla _\rho }{h^{\rho \mu }}) + ({\nabla ^\sigma }{h_{\mu \sigma }})({\nabla ^\mu }h)\nonumber\\
 &&- \frac{1}{2}({\nabla _\mu }h)({\nabla ^\mu }h) - {({h^{\mu \nu }})^2} - \frac{{d - 2}}{2}{h^2}] ,\label{2.9}
\end{eqnarray}
and cubic terms
\begin{eqnarray}
 &&L^{AdS}_{3}(h) =  - \frac{1}{2}\sqrt { - g} \{ \,h[{(\nabla h)_\mu }{(\nabla h)^\mu } - \frac{3}{2}{\nabla_{\rho}h_{\mu \nu}}{\nabla^{\nu}h^{\mu \rho}} + \frac{1}{4}{\nabla_{\rho}h_{\mu \nu}}{\nabla^{\rho}h^{\mu \nu}}\nonumber\\&&
 + \frac{1}{2}{(\nabla h)^\mu }{\nabla_{\mu}h} - \frac{1}{4}{\nabla_{\mu}h}{\nabla^{\mu}h}]
 - 2{\nabla_{\mu}h}{\nabla^{\sigma}h^{\mu \rho}}{h_{\rho \sigma }} + 2h_\nu ^\mu {\nabla^{\sigma}h^{\nu \rho}}{\nabla_{\mu}h_{\rho \sigma}}\nonumber\\ &&+ h_\nu ^\mu {\nabla_{\rho}h_{\mu \sigma}}{\nabla^{\sigma}h^{\nu \rho}} - h_\nu ^\mu {\nabla_{\sigma}h_{\mu \rho}}{\nabla^{\sigma}h^{\nu \rho}} - \frac{1}{2}h_\nu ^\mu {\nabla_{\mu}h_{\rho \sigma}}{\nabla^{\nu}h^{\rho \sigma}}
+ \frac{1}{2}h_\nu ^\mu {\nabla_{\mu}h}{\nabla^{\nu}h} \nonumber\\&&-{(\nabla h)^\mu }{\nabla_{\mu}h_{\rho \sigma}}{h^{\rho \sigma }} + {\nabla^{\mu}h}{\nabla_{\mu}h_{\rho \sigma}}{h^{\rho \sigma }} - \frac{{d+6}}{6}{h^3} + (2d+1)h{h_{\alpha \beta }}{h^{\alpha \beta }} - \frac{{4d}}{3}h_\alpha ^\beta h_\beta ^\gamma h_\gamma ^\alpha\} ,\nonumber\\ \label{2.10}
\end{eqnarray}
of a linearized gravitational Lagrangian in an $AdS$ background.

We see that after mnemonically replacing ordinary $d+2$ dimensional derivatives by $d+1$ dimensional $AdS$ covariant derivatives, and multiplying by $\sqrt{-g}$, $AdS$ expressions  (\ref{2.9}), (\ref{2.10}) differ from the flat partners (\ref{2.4}) and (\ref{2.5}) by  corresponding $1/L^{2}$ corrections
\begin{eqnarray}
  L^{AdS}_{(2)Corrections} (h)&=&- \frac{1}{2}\sqrt {- g}[ - ({h^{\mu \nu }})^2 - \frac{{d - 2}}{2}{h^2}] ,\label{2.11}\\
  L^{AdS}_{(3)Corrections}(h) &=&- \frac{1}{2}\sqrt {- g}[ - \frac{{d + 6}}{6}{h^3} + (2d + 1)h{h_{\alpha \beta }}{h^{\alpha \beta }} - \frac{{4d}}{3}h_\alpha ^\beta h_\beta ^\gamma h_\gamma ^\alpha] .\quad\quad\label{2.12}
\end{eqnarray}
These we should try to derive using the modified radial reduction scheme described in the previous section.
Actually we should for this interacting case check the following prescription:

 \emph{Writing (\ref{2.4}) and (\ref{2.5}) in curvilinear coordinates  with metric (\ref{1.13}) and Christoffel symbols   (\ref{Christ1}), (\ref{Christ2}), we should finally obtain (\ref{2.9}) and (\ref{2.10}) making use for the case $s=2$ of the overall infinite factor (\ref{1.49}) and the reduction ansatz (\ref{1.19})}.

 Taking into account that we have already a proof for the free quadratic theory of a general spin $s$, the quadratic reduction for the spin 2 case can teach us how the different order of derivatives and a careful partial integration in the Lagrangian before and after the reduction ansatz in the presence of the overall infinite factor (\ref{1.49}) may enable us to derive the corrections for the quadratic and cubic Lagrangians for any spin.

 Note also that the factor (\ref{1.49}) and the reduction ansatz (\ref{1.19}) for the spin 2 case are
 \begin{eqnarray}
   &&\left[\int du e^{du}\right] , \label{2.13}\\
   &&h_{\mu\nu}(u,x^{\mu})=e^{2u}h_{\mu\nu}(x) ,\label{2.14}\\
   &&h_{u\mu}(u,x^{\mu})=h_{uu}(u,x^{\mu})=0 .\label{2.15}
 \end{eqnarray}

 After some long but straightforward calculations we arrive at the following interesting result:
 In both free and interacting cases the radial reduction procedure exactly produces right $AdS_{d+1}$
 corrections (\ref{2.11}) and (\ref{2.12}) if we apply the reduction ansatz (\ref{2.14}) only after partial integrations of all terms with full $\partial_{u}$ derivatives. These terms all arise in the following form
 \begin{equation}\label{2.16}
  \int du e^{du}\partial_{u} {\cal{L}}(e^{-2u}h(u,x)) ,
 \end{equation}
 where ${\cal{L}}$ is a quadratic or a cubic $1/L^{2}$ correction term.
 This partial integration gives correctly all terms in (\ref{2.11}) and (\ref{2.12}) proportional to $d$.
 All other terms of (\ref{2.9}), (\ref{2.10}) are produced automatically in the right way independent of derivative ordering.
 To avoid this discrepancy with full $\partial_{u}$ derivatives we can slightly modify our reduction scheme formulating it in two steps. \emph{First we can derive the equation of motion in a flat background and apply the reduction ansatz without any freedom in derivative ordering and partial integration. Then we can integrate the resulting equation without factorized infinite factor in the $d+1$ dimensional Lagrangian.} The result coincides with the action obtained with our prescription for full $u$ derivatives.

\section{Cubic interactions of HS fields and Radial Reduction to $AdS_{d+1}$}
\setcounter{equation}{0}
\quad Now we start to consider the radial reduction in the interacting case for general spin $s$. For that we repeat the general formula for a covariant cubic interaction of higher spin gauge fields in a flat background as presented in \cite{MMR1} and \cite{MMR2}.
The main result of \cite{MMR1, MMR2} is the following. The gauge invariance fixes in a unique way the cubic interaction if the main cyclic ansatz term without divergences and traces is given. Accordingly in this article we consider only the main term of the cubic interaction postponing the proof for all other terms to a future publication, and understanding intuitively that gauge invariance is going to regulate in a correct fashion the radial reduction for all other terms presented in \cite{MMR1, MMR2} and classified in corresponding tables there.

In \cite{MMR1, MMR2} we considered three potentials $h^{(s_{1})}(X_{1};a^{A}), h^{(s_{2})}(X_{2};b^{A}), h^{(s_{3})}(X_{3};c^{A})$ of $d+2$ dimensional flat theory with ordered  spins $s_{i}$
\begin{equation}
s_{1} \geq s_{2}\geq s_{3} ,\label{3.1}
\end{equation}
and with the cyclic ansatz for the interaction
\begin{eqnarray}
&&\mathcal{L}_{I}^{main}(h^{(s_{1})}(X_{1},a^{A}),h^{(s_{2})}(X_{2},b^{A}),h^{(s_{3})}(X_{3}c^{A}))\nonumber\\
&&\hspace{2cm}=\sum_{n_{i}} C_{n_{1},n_{2},n_{3}}^{s_{1},s_{2},s_{3}} \int d^{d+2}X_{1}d^{d+2}X_{2}d^{d+2}X_{3} \delta (X_{3}-X_{1}) \delta(X_{2}-X_{1})\nonumber\\
&&\hspace{2cm}\times\tilde{T}(Q_{12},Q_{23},Q_{31}|n_{1},n_{2},n_{3})h^{(s_{1})}
(X_{1};a^{A})h^{(s_{2})}(X_{2};b^{B})h^{(s_{3})}(X_{3};c^{C}) ,\nonumber\\\label{3.2}
\end{eqnarray}
where
\begin{eqnarray}
&&\tilde{T}(Q_{12},Q_{23},Q_{31}|n_{1},n_{2},n_{3})\nonumber\\ &&\hspace{2cm}=(\partial_{a^{A}}\partial_{b_{A}})^{Q_{12}}(\partial_{b^{B}}\partial_{c_{B}})^{Q_{23}} (\partial_{c^{C}}\partial_{a_{C}})^{Q_{31}}(\partial_{a^{D}}\tilde{\nabla}_{2}^{D})^{n_{1}}(\partial_{b^{E}}\tilde{\nabla}^{E}_{3})^{n_{2}}( \partial_{c^{F}}\tilde{\nabla}^{F}_{1})^{n_{3}} ,\nonumber\\\label{3.3}
\end{eqnarray}
and the notation $"main"$ as a superscript means that it is an ansatz for terms without $Divh^{(s_{i}-1)}$ and $Trh^{(s_{i}-2)}$.
Denoting the number of derivatives by $\Delta$ we have
\begin{equation}
n_{1}+n_{2}+n_{3} = \Delta . \label{3.4}
\end{equation}
We shall later determine and then use the minimal possible $\Delta$. As balance equations we have
\begin{eqnarray}
n_{1}+Q_{12}+Q_{31} = s_{1} , \nonumber\\
n_{2}+Q_{23}+Q_{12} = s_{2} ,\nonumber\\
n_{3} + Q_{31} + Q_{23} = s_{3} .\label{3.5}
\end{eqnarray}
These equations are solved by
\begin{eqnarray}
Q_{12} = n_{3}-\nu_{3} ,\nonumber\\
Q_{23} = n_{1} - \nu_{1} , \nonumber\\
Q_{31} = n_{2} - \nu_{2} .\label{3.6}
\end{eqnarray}
Since the l.h.s. cannot be negative, we have
\begin{equation}
n_{i} \geq  \nu_{i} .\nonumber
\end{equation}
The $\nu_{i}$ are determined to be
\begin{equation}
\nu_{i} = 1/2 (\Delta +s_{i} -s_{j} -s_{k}), \quad i,j,k \quad  \textnormal{are all different.}\label{3.7}
\end{equation}
It follows that the minimally possible $\Delta$ is expressed by Metsaev's  \cite{Metsaev} (using the ordering of the $s_{i}$).
\begin{equation}
\Delta_{min} = \max{[s_{i} +s_{j} -s_{k}]} = s_{1}+ s_{2} -s_{3} .\label{3.8}
\end{equation}
Another result of \cite{MMR1, MMR2} is the trinomial expression for the coefficients in (\ref{3.2}) fixed by Noether's procedure. Taking into account (\ref{3.5})-(\ref{3.8}) we can write it in the following elegant form
\begin{eqnarray}
  C_{n_{1},n_{2},n_{3}}^{s_{1},s_{2},s_{3}}&=& C_{Q_{12},Q_{23},Q_{31}}^{s_{1},s_{2},s_{3}} = const\quad {s_{min} \choose Q_{12},Q_{23},Q_{31}} .\label{3.9}
\end{eqnarray}

To investigate a possibility of application for radial reduction in the case of the cubic interaction, we rewrite (\ref{3.2}), (\ref{3.3}) in another form more convenient for the reduction
\begin{eqnarray}
&&\mathcal{L}_{I}^{main}(h^{(s_{1})}(X,a^{A}),h^{(s_{2})}(X,b^{A}),h^{(s_{3})}(X,c^{A}))
=\nonumber\\ &&\sum_{Q_{ij}} C_{Q_{12},Q_{23},Q_{31}}^{s_{1},s_{2},s_{3}}\int d^{d+2}X *^{Q_{31}+n_{3}}_{c^{A}}K^{(s_{1})}(Q_{31},n_{3};c^{A},a^{A};X)\nonumber\\
&&*^{Q_{12}+n_{1}}_{a^{A}}K^{(s_{2})}(Q_{12},n_{1};a^{A},b^{A};X)*^{Q_{23}+n_{2}}_{b^{A}}
K^{(s_{3})}(Q_{23},n_{2};b^{A},c^{A};X) ,
\quad\quad\quad\nonumber\\\label{3.10}
\end{eqnarray}
where
\begin{eqnarray}
&&K^{(s_{1})}(Q_{12},n_{1};a^{A},b^{A};X)=(a^{A}\partial_{b^{A}})^{Q_{12}}
(a^{B}\tilde{\nabla}_{B})^{n_{1}}h^{(s_{1})}(X;b^{C}) .\label{3.11}
\end{eqnarray}
So we see that we can express our cubic interaction as a cube of a bitensor function
\begin{equation}\label{3.12}
    K^{(s)}(Q,n;\tilde{a},\tilde{b};X)=(\tilde{a}\partial_{\tilde{b}})^{Q}
(\tilde{a}\tilde{\nabla})^{n}h^{(s)}(X;\tilde{b}) ,
\end{equation}
with cyclic index contraction.
From (\ref{3.12}) we use for shortness the notation $\tilde{a}, \tilde{b}, \dots$ for $d+2$ dimensional objects in curvilinear coordinates to distinguish them from $AdS_{d+1}$ dimensional objects when we have index summation. In other words
\begin{eqnarray}
  (\tilde{a}\partial_{\tilde{b}}) &=& a^{A}\partial_{b^{A}}=a^{u}\partial_{b^{u}}+(a\partial_{b}) ,\label{3.13}\\
  (a\partial_{b})&=&a^{\mu}\partial_{b^{\mu}} ,\label{3.14}
\end{eqnarray}
and
\begin{eqnarray}
  (\tilde{a}\tilde{\nabla}) &=& a^{B}\tilde{\nabla}_{B}=a^{u}\nabla_{u}+(a D) ,\label{3.15}\\
  (aD)&=&a^{\mu}D_{\mu} .\label{3.16}
\end{eqnarray}

Another important point here is the difference in definition of the covariant differentiation operators
(\ref{1.15}) and (\ref{1.16}) in the case of interaction. The minimal object here is a bitensor (\ref{3.12}) which has two sets of symmetrized indices (i.e. $a$ and $b$ in (\ref{3.12})). In this case we should define covariant differentiation operators for both sets of indices:
\begin{eqnarray}
\nabla_{u}&=& \partial_{u}-a^{u}\partial_{a^{u}}-a^{\mu}\partial_{a^{\mu}}-b^{u}\partial_{b^{u}}-b^{\mu}\partial_{b^{\mu}} ,\label{3.17}\\
    D_{\mu} &=& \nabla_{\mu}-a^{u}\partial_{a^{\mu}}-a_{\mu}\partial_{a^{u}}-b^{u}\partial_{b^{\mu}}-b_{\mu}\partial_{b^{u}} .\label{3.18}
  \end{eqnarray}
Now we have all ingredients to start analyzing the scaling behaviour of interaction Lagrangian (\ref{3.10}) in curvilinear coordinates (\ref{1.13}). First of all we note that in the new frame the measure and the star contractions only create an additional $u$ phase
\begin{eqnarray}
  \int d^{d+2}X &=& \int du e^{(d+2)u}\sqrt{g} ,\label{3.19}\\
  *^{Q_{12}+n_{1}}_{\tilde{a}}&=& e^{-2(Q_{12}+n_{1})u}\prod^{Q_{12}+n_{1}}_{k=1}(
  \overleftarrow{\partial}_{a^{u_{k}}}\overrightarrow{\partial}_{a^{u_{k}}}
  +\overleftarrow{\partial}_{a^{\mu_{k}}}g^{\mu_{k}\nu_{k}}\overrightarrow{\partial}_{a^{\nu_{k}}}) ,\label{3.20}\\
  *^{Q_{23}+n_{2}}_{\tilde{b}}&=& e^{-2(Q_{23}+n_{2})u}\prod^{Q_{23}+n_{2}}_{k=1}(
  \overleftarrow{\partial}_{b^{u_{k}}}\overrightarrow{\partial}_{b^{u_{k}}}
  +\overleftarrow{\partial}_{b^{\mu_{k}}}g^{\mu_{k}\nu_{k}}\overrightarrow{\partial}_{b^{\nu_{k}}}) ,\label{3.21}\\
  *^{Q_{31}+n_{3}}_{\tilde{c}}&=& e^{-2(Q_{31}+n_{3})u}\prod^{Q_{31}+n_{3}}_{k=1}(
  \overleftarrow{\partial}_{c^{u_{k}}}\overrightarrow{\partial}_{c^{u_{k}}}
  +\overleftarrow{\partial}_{c^{\mu_{k}}}g^{\mu_{k}\nu_{k}}\overrightarrow{\partial}_{c^{\nu_{k}}}) .\label{3.22}
\end{eqnarray}
Then taking into account our "$u$" dependence restriction (\ref{1.19}) for all three different spin fields, we obtain an additional factor
\begin{equation}\label{3.23}
    e^{[2(s_{1}-1)+2(s_{2}-1)+2(s_{3}-1)]u} ,
\end{equation}
supplemented with the substitution
\begin{equation}\label{3.24}
    \partial_{u}=> 2(s_{i}-1) ,
\end{equation}
in corresponding differentiations, we arrive at the final overall factor
\begin{equation}\label{3.25}
    e^{[d-4-2(\sum Q_{ij}+\sum n_{i}-\sum s_{i})]u} .
\end{equation}
Summing lines in (\ref{3.5}) and using the definition for the number of derivatives (\ref{3.4}) we obtain the following expression
\begin{eqnarray}
&& e^{[\sum s_{i}-\Delta+d-4]u} .\label{3.26}
\end{eqnarray}
So we see now that only in the case of the minimal numbers of derivatives (\ref{3.8}) this exponent coincides with the infinite factor in front of the free action for the gauge field with minimal spin (\ref{1.47}). In the case of three spins ordered as $s_{1}\geq s_{2}\geq s_{3}$ it is
\begin{eqnarray}
   \sum s_{i}- \Delta_{min}+d-4 &=&d+2s_{3}-4 ,\nonumber\\
\end{eqnarray}
with the obvious limit $d+2s-4$ in the self-interacting case $s_{1}=s_{2}=s_{3}=s$. So we see that the cubic interaction in the case of the minimal number of derivatives is relevant for the radial reduction procedure described in the previous section. Therefore it should produce the right $1/L^{2}$ corrections for the main term of the cubic interaction in $AdS_{d+1}$.

\section{First $AdS$ corrections in the self-interacting case}
\setcounter{equation}{0}
The main term in the case of a cubic self-interaction can be obtained from the general expressions (\ref{3.2})-(\ref{3.6}) taking
\begin{eqnarray}
  s_{1} &=& s_{2}=s_{3}=s ,\\
  \nu_{1} &=& \nu_{2}=\nu_{3}=0 ,\\
   Q_{23} &=& n_{1}=\alpha ,\\
   Q_{31} &=& n_{2}=\beta ,\\
   Q_{12} &=& n_{3}=\gamma .
\end{eqnarray}
Then (\ref{3.2}), (\ref{3.3}) in curvilinear coordinates (\ref{1.8})-(\ref{1.12}) transform to
\begin{eqnarray}
&&\mathcal{L}_{I}^{main}=\sum_{\alpha, \beta, \gamma \atop \alpha+\beta+\gamma=s} {s \choose \alpha,\beta,\gamma} \int d^{d+2}X\sqrt{G}
(\partial_{a^{A}}\partial_{b_{A}})^{\gamma}(\partial_{b^{B}}\partial_{c_{B}})^{\alpha} (\partial_{c^{C}}\partial_{a_{C}})^{\beta}\nonumber \\
&&\left[(\partial_{c^{F}}\tilde{\nabla}^{F})^{\gamma}h^{(s)}(X;a^{A})\right]
\left[(\partial_{a^{D}}\tilde{\nabla}^{D})^{\alpha}h^{(s)}(X;b^{B})\right]
\left[(\partial_{b^{E}}\tilde{\nabla}^{E})^{\beta}h^{(s)}(X;c^{C})\right] ,\label{6}
\end{eqnarray}
where space-time covariant derivatives inside of each rectangular bracket in the second line work with the corresponding field inside of the bracket but contraction derivatives $\partial_{a}, \partial_{b}, \partial_{c}$ act on the corresponding field $h^{(s)}(X;a), h^{(s)}(X;b)$ or $h^{(s)}(X;c)$ in a cyclic way. Then we see that $1/L^{2}$ corrections can arise in two different ways from the $\partial_{a^{u}}$ and $\nabla_{u}$ differentiations. The first source is the expansion of terms in the second line of  (\ref{6}). Using (\ref{1.14})-(\ref{1.16}) we have
\begin{eqnarray}\label{7}
  (\partial_{c^{F}}\tilde{\nabla}^{F})^{\gamma}=e^{-2\gamma u}(-\partial_{c^{u}}\nabla_{u}
  +g^{\mu\nu}\partial_{c^{\mu}}D_{\nu})^{\gamma}\nonumber\\
  =e^{-2\gamma u}[(\partial_{c^{\mu}}D^{\mu})^{\gamma}
  -\gamma\partial_{c^{u}}\nabla_{u}(\partial_{c^{\mu}}D^{\mu})^{\gamma-1}+\dots] .
\end{eqnarray}
Using this expansion in powers of $(L^2)^{-1}$ for all three rectangular brackets in the second line of (\ref{6}) we obtain correction terms with full $\partial_{u}$ derivatives similar to the $s=2$ case in section 2
\begin{eqnarray}
&&\mathcal{L}_{1}^{1/L^{2}}=s(s-1)\sum_{\alpha, \beta, \gamma \atop \alpha+\beta+\gamma=s-2} {s-2 \choose \alpha,\beta,\gamma} \int d^{d+1}x \sqrt{g} du e^{(d+2-s)u}
(\partial_{a}\partial_{b})^{\gamma+1}(\partial_{b}\partial_{c})^{\alpha+1} (\partial_{c}\partial_{a})^{\beta+1}\nonumber \\
&&\partial_{u}\left\{\left[(\partial_{c}\nabla)^{\gamma}e^{-su}h^{(s)}(u,x;a)\right]
\left[(\partial_{a}\nabla)^{\alpha}e^{-su}h^{(s)}(u,x;b)\right]
\left[(\partial_{b}\nabla)^{\beta}e^{-su}h^{(s)}(u,x;c)\right]\right\} ,\label{8}
\end{eqnarray}
where we used the notations
\begin{eqnarray}
  &&(\partial_{a}\partial_{b})=g^{\mu\nu}\partial_{a^{\mu}}\partial_{b^{\nu}}=(\partial_{a^{\mu}}\partial_{b_{\mu}}), \dots ,\label{x}\\
  &&(\partial_{a}\nabla)=g^{\mu\nu}\partial_{a^{\mu}}\nabla^{AdS}_{\nu}
  =(\partial_{a_{\mu}}\nabla_{\mu}), \dots ,\label{y}
\end{eqnarray}
 and discovered the first step of our radial reduction putting to zero all $u$ components of the tensor gauge field $h^{(s)}(u,x;a)=h^{(s)}(u,x^{\mu};a^{\mu})$ at the end of the calculations. Following the prescription of section 2 we performed a partial integration in direction $u$ and then only applied the final ansatz for the $u$ dependence of the $h^{(s)}$ field in $AdS$ space (\ref{1.19})
\begin{equation}\label{9}
  e^{-su}h^{(s)}(u,x;a)=e^{(s-2)u}h^{(s)}(x;a) .
\end{equation}
Thus we arrived at the final expression for this $1/L^{2}$ correction term
\begin{eqnarray}
&&\mathcal{L}_{1}^{1/L^{2}}=\left[\int du e^{(d+2s-4)u}\right] s(s-1)(s-d-2)\sum_{\alpha, \beta, \gamma \atop \alpha+\beta+\gamma=s-2} {s-2 \choose \alpha,\beta,\gamma}\int d^{d+1}x \sqrt{g}
\nonumber \\
&&(\partial_{a}\partial_{b})^{\gamma+1}(\partial_{b}\partial_{c})^{\alpha+1} (\partial_{c}\partial_{a})^{\beta+1}\left[(\partial_{c}\nabla)^{\gamma}h^{(s)}(x;a)\right]
\left[(\partial_{a}\nabla)^{\alpha}h^{(s)}(x;b)\right]
\left[(\partial_{b}\nabla)^{\beta}h^{(s)}(x;c)\right] ,\nonumber\\\label{9(1)}
\end{eqnarray}
with the same infinite factor in front as before.

The second source for a correction arises in the expansion of three contraction terms like
\begin{eqnarray}\label{10}
  (\partial_{a^{A}}\partial_{b_{A}})^{\gamma}=e^{-2\gamma u}(-\partial_{a^{u}}\partial_{b^{u}}
  +g^{\mu\nu}\partial_{a^{\mu}}\partial_{b^{\nu}})^{\gamma} \nonumber\\
  =e^{-2\gamma u}[(\partial_{a}\partial_{b})^{\gamma}-\gamma\partial_{a^{u}}\partial_{b^{u}} (\partial_{a}\partial_{b})^{\gamma-1}+\dots] .
\end{eqnarray}
Performing a calculation similar to the previous case we obtain a final expression for this correction:
\begin{eqnarray}
&&\mathcal{L}_{2}^{1/L^{2}}=-\left[\int du e^{(d+2s-4)u}\right] (s+1)s(s-1)\sum_{\alpha, \beta, \gamma \atop \alpha+\beta+\gamma=s-2} {s-2 \choose \alpha,\beta,\gamma}\int d^{d+1}x \sqrt{g}
\nonumber \\
&&(\partial_{a}\partial_{b})^{\gamma+1}(\partial_{b}\partial_{c})^{\alpha+1} (\partial_{c}\partial_{a})^{\beta+1}\left[(\partial_{c}\nabla)^{\gamma}h^{(s)}(x;a)\right]
\left[(\partial_{a}\nabla)^{\alpha}h^{(s)}(x;b)\right]
\left[(\partial_{b}\nabla)^{\beta}h^{(s)}(x;c)\right] .\nonumber\\\label{11}
\end{eqnarray}
Note that the second correction looks similar to the first one but without $d$ dependent factor coming from the partial $u$ integration which we do not have in the second case. The overall factor $s+1$ specific for the second case arose from the combination
$(\alpha+\beta+\gamma+3)$ surviving in front of the trinomial reduced to the level $\alpha+\beta+\gamma=s-2$.

\section{All order $AdS$ corrections for the main term of the self-interaction}
\setcounter{equation}{0}
Led by the success of the considerations in the previous section we start to calculate all $1/L^{2}$ corrections for the main term (\ref{6}). For that purpose we expand all three contracting terms till the end in (\ref{6})
\begin{eqnarray}
&&(\partial_{a^{A}}\partial_{b_{A}})^{\gamma}(\partial_{b^{B}}\partial_{c_{B}})^{\alpha} (\partial_{c^{C}}\partial_{a_{C}})^{\beta}=e^{-2su}\sum_{n_{3}=0}^{\gamma}\sum_{n_{2}=0}^{\beta}\sum_{n_{1}=0}^{\alpha}
\binom{\gamma}{n_{3}}\binom{\alpha}{n_{1}}\binom{\beta}{n_{2}}(-1)^{n_{1}+n_{2}+n_{3}}\nonumber\\&&
(\partial_{a^{\mu}}\partial_{b_{\mu}})^{\gamma-n_{3}}
(\partial_{b^{\nu}}\partial_{c_{\nu}})^{\alpha-n_{1}}(\partial_{c^{\lambda}}\partial_{a_{\lambda}})^{\beta-n_{2}} (\partial_{a^{u}}\partial_{b^{u}})^{n_{3}}(\partial_{b^{u}}\partial_{c^{u}})^{n_{1}}
(\partial_{c^{u}}\partial_{a^{u}})^{n_{2}} ,\label{a1}
\end{eqnarray}
and do the same with all three covariant derivative terms
\begin{eqnarray}
  &&\left[(\partial_{c^{F}}\tilde{\nabla}^{F})^{\gamma}h^{(s)}(X;a^{A})\right]
\left[(\partial_{a^{D}}\tilde{\nabla}^{D})^{\alpha}h^{(s)}(X;b^{B})\right]
\left[(\partial_{b^{E}}\tilde{\nabla}^{E})^{\beta}h^{(s)}(X;c^{C})\right]\nonumber\\
&&=e^{-2su}\sum_{m_{3}=0}^{\gamma}\sum_{m_{2}=0}^{\beta}\sum_{m_{1}=0}^{\alpha}
\binom{\gamma}{m_{3}}\binom{\alpha}{m_{1}}\binom{\beta}{m_{2}}(-1)^{m_{1}+m_{2}+m_{3}}\nonumber\\
&&\left[(\partial_{c^{\mu}}D^{\mu})^{\gamma-m_{3}}(\partial_{c^{u}}\nabla_{u})^{m_{3}}h^{(s)}(u,x^{\mu};a^{u},a^{\mu})\right]
\left[(\partial_{a^{\nu}}D^{\nu})^{\alpha-m_{1}}(\partial_{a^{u}}\nabla_{u})^{m_{1}}h^{(s)}(u,x^{\mu};b^{u},b^{\mu})\right]\nonumber\\
&&\left[(\partial_{b^{\lambda}}D^{\lambda})^{\beta-m_{2}}
(\partial_{b^{u}}\nabla_{u})^{m_{2}}h^{(s)}(u,x^{\mu};c^{u},c^{\mu})\right] .\label{a2}
\end{eqnarray}
Then collecting the same type $\partial_{a^{u}}, \partial_{b^{u}}, \partial_{c^{u}}$ of derivatives from both expressions (\ref{a1}) and (\ref{a2}), we can perform the first step of our reduction procedure putting to zero all $u$ components of the final expressions to zero after the differentiations and
contractions in a careful way.  For that we should use the following crucial formulas
\begin{eqnarray}
 &&\left[(\partial_{a^{u}})^{m_{1}+n_{2}+n_{3}}
 (\partial_{c^{\mu}}D^{\mu})^{\gamma-m_{3}}h^{(s)}(u,x^{\mu};a^{u},a^{\mu})\right]|_{a^{u}=0} =\frac{(\gamma-m_{3})!(-1)^{m_{1}+n_{2}+n_{3}}}{(\gamma-m_{3}-m_{1}-n_{2}-n_{3})!}\nonumber\\&& (\partial_{a^{\nu}}\partial_{c_{\nu}})^{m_{1}+n_{2}+n_{3}}
 (\partial_{c^{\mu}}\nabla^{\mu})^{\gamma-m_{3}-m_{1}-n_{2}-n_{3}}h^{(s)}(u,x^{\mu};a^{\mu})\quad + \textit{trace terms} ,\label{a3}\\
 &&\left[(\partial_{b^{u}})^{m_{2}+n_{1}+n_{3}}
 (\partial_{a^{\mu}}D^{\mu})^{\alpha-m_{1}}h^{(s)}(u,x^{\mu};b^{u},b^{\mu})\right]|_{b^{u}=0} =\frac{(\alpha-m_{1})!(-1)^{m_{2}+n_{1}+n_{3}}}{(\alpha-m_{1}-m_{2}-n_{1}-n_{3})!}\nonumber\\&& (\partial_{a^{\nu}}\partial_{b_{\nu}})^{m_{2}+n_{1}+n_{3}}
 (\partial_{c^{\mu}}\nabla^{\mu})^{\alpha-m_{1}-m_{2}-n_{1}-n_{3}}h^{(s)}(u,x^{\mu};b^{\mu})\quad + \textit{trace terms} ,\label{a4}\\
 &&\left[(\partial_{c^{u}})^{m_{3}+n_{1}+n_{2}}
 (\partial_{b^{\mu}}D^{\mu})^{\beta-m_{2}}h^{(s)}(u,x^{\mu};c^{u},c^{\mu})\right]|_{c^{u}=0} =\frac{(\beta-m_{2})!(-1)^{m_{3}+n_{1}+n_{2}}}{(\beta-m_{2}-m_{3}-n_{1}-n_{2})!}\nonumber\\&& (\partial_{b^{\nu}}\partial_{c_{\nu}})^{m_{3}+n_{1}+n_{2}}
 (\partial_{b^{\mu}}\nabla^{\mu})^{\beta-m_{2}-m_{3}-n_{1}-n_{2}}h^{(s)}(u,x^{\mu};c^{\mu})\quad + \textit{trace terms}.\label{a5}
\end{eqnarray}
After that we take into account that $\alpha+\beta+\gamma=s$ and rearrange the summations coming from (\ref{a1}), (\ref{a2}) in the following way
\begin{eqnarray}
 \sum_{n_{3}\geq 0}\sum_{n_{2}\geq 0}\sum_{n_{1}\geq 0} (-1)^{n_{1}+n_{2}+n_{3}}&=&\sum_{N\geq 0}(-1)^{N}\sum_{{n_{1}, n_{2}, n_{3}\atop n_{1}+n_{2}+n_{3}=N}} ,\label{a6}\\
  \sum_{m_{3}\geq 0}\sum_{m_{2}\geq 0}\sum_{m_{1}\geq 0} (-1)^{m_{1}+m_{2}+m_{3}}&=&\sum_{M \geq 0}(-1)^{M}\sum_{{n_{1}, n_{2}, n_{3}\atop m_{1}+m_{2}+m_{3}=M}} .\label{a7}
\end{eqnarray}
After that we introduce instead of $\alpha, \beta, \gamma$ new summation variables
\begin{eqnarray}
  \tilde{\alpha} &=&\alpha-m_{1}-m_{2}-n_{1}-n_{3}=\alpha^{\ast}(m)-m_{1}-m_{2} ,\label{a8}\\
  \tilde{\beta}&=&\beta-m_{2}-m_{3}-n_{1}-n_{2}=\beta^{\ast}(m)-m_{2}-m_{3} ,\label{a9}\\
  \tilde{\gamma}&=&\gamma-m_{3}-m_{1}-n_{2}-n_{3}=\gamma^{\ast}(m)-m_{3}-m_{1} .\label{a10}
\end{eqnarray}
 with corresponding summation limits and constraints
 \begin{eqnarray}
 &&0\leq \tilde{\alpha}, \tilde{\beta}, \tilde{\gamma} \leq s-2(M+N) ,\label{a11}\\
 && \tilde{\alpha}+\tilde{\beta}+\tilde{\gamma}=s-2(M+N) ,\label{a12}\\
 && \alpha^{\ast}(m)+\beta^{\ast}(m)+\gamma^{\ast}(m)=s-2N ,\label{a13}\\
 && M=n_{1}+n_{2}+n_{3} ,\label{a14}\\
 && N=m_{1}+m_{2}+m_{3} .\label{a15}
 \end{eqnarray}
Collecting all terms, using the shortening notations (\ref{x}), (\ref{y}) for the contraction of $AdS$ indices, and taking relations $(\nabla_{u})^{m_{i}}h^{(s)}(u)=e^{su}(\partial_{u})^{m_{i}} e^{-su}h^{(s)}(u)$ into account, we arrive at the following important preliminary expression
\begin{eqnarray}
&&\int du e^{(d+2-s)u} \int d^{d+1}x\sqrt{g} \sum_{N}\sum_{M}\frac{(-1)^{N}s!}{(s-2(N+M))!}
\sum_{{m_{1}, m_{2}, m_{3}\atop m_{1}+m_{2}+m_{3}=M}}\frac{1}{m_{1}!m_{2}!m_{3}!}\nonumber\\
&& \sum_{{n_{1}, n_{2}, n_{3}\atop n_{1}+n_{2}+n_{3}=N}}\binom{\alpha^{\ast}(m)+n_{1}+n_{3}}{n_{1}}
\binom{\beta^{\ast}(m)+n_{1}+n_{2}}{n_{2}}\binom{\gamma^{\ast}(m)+n_{3}+n_{2}}{n_{3}}\nonumber\\
&&\sum_{{\tilde{\alpha},\tilde{\beta},\tilde{\gamma}\atop \tilde{\alpha}+\tilde{\beta}+\tilde{\gamma}= s-2(N+M)}}
{s-2(N+M) \choose \tilde{\alpha},\tilde{\beta},\tilde{\gamma}}(\partial_{a}\partial_{b})^{\tilde{\gamma}+N+M}
(\partial_{b}\partial_{c})^{\tilde{\alpha}+N+M}(\partial_{c}\partial_{a})^{\tilde{\beta}+N+M}\nonumber\\
&&\left[(\partial_{u})^{m_{3}}(\partial_{c}\nabla)^{\tilde{\gamma}}e^{-su}h^{(s)}(x;a)\right]
\left[(\partial_{u})^{m_{1}}(\partial_{a}\nabla)^{\tilde{\alpha}}e^{-su}h^{(s)}(x;b)\right]
\left[(\partial_{u})^{m_{2}}(\partial_{b}\nabla)^{\tilde{\beta}}e^{-su}h^{(s)}(x;c)\right].\nonumber\\ \label{a16}
\end{eqnarray}
For finishing our task we should investigate the sum in the second line of this formula. Using a computer we learned that the result is a symmetric
polynomial  $P_{N}(\alpha^{\ast}, \beta^{\ast}, \gamma^{\ast})$  in $ \alpha^{\ast}(m), \beta^{\ast}(m), \gamma^{\ast}(m)$ defined as
\begin{equation}\label{a17}
  \frac{1}{N!}P_{N}(\alpha^{\ast}, \beta^{\ast}, \gamma^{\ast})=\sum_{{n_{1}, n_{2}, n_{3}\atop n_{1}+n_{2}+n_{3}=N}}\binom{\alpha^{\ast}(m)+n_{1}+n_{3}}{n_{1}}
\binom{\beta^{\ast}(m)+n_{1}+n_{2}}{n_{2}}\binom{\gamma^{\ast}(m)+n_{3}+n_{2}}{n_{3}} ,
\end{equation}
that depends only on the sum of variables
\begin{eqnarray}
  &&\alpha^{\ast}(m)+\beta^{\ast}(m)+\gamma^{\ast}(m)=s-2N ,\label{a18}\\
  && P_{N}(\alpha^{\ast}, \beta^{\ast}, \gamma^{\ast})=P_{N}(s-2N) ,\label{a19}
  \end{eqnarray}
and therefore  the second line of (\ref{a16}) does not depend on $m_{1}, m_{2}, m_{3}$.
As a result we can take the sum over $m_{1}, m_{2}, m_{3}$ and obtain the full $u$ derivative of power $M=n_{1}+n_{2}+n_{3}$.

Performing $M$ times a partial integration over $u$ and restricting the $u$ dependence of the field after that in agreement with our second step of the radial reduction (\ref{9}) we obtain finally the following elegant formula
\begin{eqnarray}
&&\int du e^{(d+2s-4)u} \int d^{d+1}x\sqrt{g}\sum_{N}\sum_{M}\frac{(-1)^{N+M}s!(d+2-s)^{M}P_{N}(s-2N)}{(s-2(N+M))!M!N!}
\nonumber\\ &&\sum_{{\tilde{\alpha},\tilde{\beta},\tilde{\gamma}\atop \tilde{\alpha}+\tilde{\beta}+\tilde{\gamma}= s-2(N+M)}}
{s-2(N+M) \choose \tilde{\alpha},\tilde{\beta},\tilde{\gamma}}(\partial_{a}\partial_{b})^{\tilde{\gamma}+N+M}
(\partial_{b}\partial_{c})^{\tilde{\alpha}+N+M}(\partial_{c}\partial_{a})^{\tilde{\beta}+N+M} \nonumber\\ &&\left[(\partial_{c}\nabla)^{\tilde{\gamma}}h^{(s)}(x;a)\right]
\left[(\partial_{a}\nabla)^{\tilde{\alpha}}h^{(s)}(x;b)\right]
\left[(\partial_{b}\nabla)^{\tilde{\beta}}h^{(s)}(x;c)\right] ,\label{a20}
\end{eqnarray}
which inherits a structure similar to that of the main term but with diminished couples of covariant derivatives and a nontrivial numerical factor.

\section{Concluding remarks}

\quad We have constructed the $AdS$ corrections to the main term of the cubic self-interaction with the minimal number $\Delta_{min}$ of derivatives for three equal spins by a new modified method of radial reduction where all quantum fields are carried by a real  AdS space.
For given spin s and $\Delta_{min} = s$  we derived all $1/L^{2}$ correction terms (\ref{a20}) in the elegant form of series of terms with numbers $s-2(M+N)$ of derivatives, where  $0 \leq M+N \leq \frac{s}{2}$. This we proved for the main terms of the cubic interaction which do not contain trace or deDonder terms. They appear with coefficients that are polynomials in the dimension $d+1$ and spin number $s$  with rational coefficients.
From this beautiful form of the $AdS$ corrections to the main term we can expect that the same method can be used for the derivation of the $AdS$corrections to the remaining terms of the cubic interaction including traces and deDonder terms connected with the main term by Noether's  procedure described in details for the flat case in \cite{MMR1} and \cite{MMR2}.
\section*{Acknowledgments}
 \quad We are indebted to Hrachya Khachatryan for accurate derivation and careful checking of the formulas for linearized interacting Einstein-Hilbert actions in flat and $AdS$  backgrounds, which helped us a lot to find the correct way of formulating radial reduction in the general case.

 This work is partially supported by Volkswagen Foundation.
R. Manvelyan and R. Mkrtchyan were partially supported by the grant of the Science Committee of the Ministry of Science and Education of the Republic of Armenia under contract 11-1c037. Work of R. Manvelyan was also partially supported by grant ANSEF 2012.

\section*{Appendix: Differential operator algebras}
\setcounter{equation}{0}
\renewcommand{\theequation}{A.\arabic{equation}}
\quad
\subsection*{ A. The algebra generated by $a_0, a_{\mu}, \partial_{a^0}, \partial_{a^{\mu} }$  and $(a,\hat\nabla)$    }
\setcounter{equation}{0}
In this appendix we consider a possible radial reduction algorithm for the main object of equation of motion including cubic interaction (\ref{3.10}): the bitensorial function $ K^{(s)}(Q,n;\tilde{a},\tilde{b};X)$
\begin{equation}
    K^{(s)}(Q,n;\tilde{a},\tilde{b};X)=(\tilde{a}\partial_{\tilde{b}})^{Q}
(\tilde{a}\tilde{\nabla})^{n} h^{(s)}(X;\tilde{b}) .\label{4.1}
\end{equation}
This term should generate all $AdS$ corrections proportional to powers of $1/L^{2}$.
For that we study these operators in a representation that act on ground states
\begin{equation}
h^{(s)}(X;\tilde{b})|_{b^{u}=0} = h^{(s)}(u,x^{\mu};b^{\mu})=e^{2(s-1)u}h^{(s)}(x^{\mu};b^{\mu}) .\label{4.2}
\end{equation}
Then we can obtain these $AdS$ corrections expanding all flat $d+2$ dimensional objects in term of $d+1$ dimensional $AdS$ space derivatives and vectors, and contracting over all $a^{u}, b^{u}, c^{u}$, and replacing $\partial_{u}$ with $2(s-1)$.
For convenience we replace from now on all $"u"$-components of auxiliary vectors by $"0"$.

So we must deal with the following $d+1$ dimensional expansion for the $n$'th power of $d+2$ dimensional derivatives
\begin{eqnarray}\label{4.3}
  (\tilde{a}\tilde{\nabla})^{n}&=& (a^{0}\nabla_{u}+a^{\mu}D_{\mu})^{n} ,
\end{eqnarray}
where the operators
\begin{eqnarray}
\nabla_{u}&=& \partial_{u}-a^{0}\partial_{a^{0}}-a^{\mu}\partial_{a^{\mu}}-b^{0}\partial_{b^{0}}-b^{\mu}\partial_{b^{\mu}} ,\label{op1}\\
    D_{\mu} &=& \nabla_{\mu}-a^{0}\partial_{a^{\mu}}-a_{\mu}\partial_{a^{0}}-b^{0}\partial_{b^{\mu}}-b_{\mu}\partial_{b^{0}} ,\label{op2}
 \label{4.5}
  \end{eqnarray}
should be realized on ground state (\ref{4.2}). It follows immediately that we can replace $\partial_{u}$ by $2(s-1)$  and $b^{0}\partial_{b^{0}}+b^{\mu}\partial_{b^{\mu}}$ by $s$. So effectively we have
\begin{equation}
  \nabla_{u}=s-2-a^{0}\partial_{a^{0}}-a^{\mu}\partial_{a^{\mu}} ,
\label{4.6}
\end{equation}
and
\begin{eqnarray}
a^{0}\nabla_{u}+a^{\mu}D_{\mu} &=&a^{\mu}\nabla^{AdS}_{\mu}(g)-a^{\mu}R_{\mu}^{b}-R ,\\
R_{\mu}^{b} &=& b^{0}\partial_{b^{\mu}} + b_{\mu}\partial_{b^{0}},\\
(a,R^{b}) &=&(a,b)\partial_{b^{0}} + b^{0}(a,\partial_{b}),\\
 R &=& a^{0}[2(a \partial_{a})-s+2]+(a^2+(a^{0})^{2})\partial _{a^{0}},
\label{4.10}
\end{eqnarray}
act on ground states (\ref{4.2}). These ground states can be characterized by the fact that
they are annihilated by the operator $\partial_{b^{0}}$, and by the total symmetry in the argument.

The operator of interest is
\begin{equation}
((a,\hat{\nabla})- R)^{n},
\label{4.11}
\end{equation}
where
\begin{equation}
    \hat{\nabla}_{\mu}=\nabla^{AdS}_{\mu}(g)-R_{\mu}^{b}.
\label{4.12}
\end{equation}
It is advantageous in the sequel to write the operator (\ref{4.10})
in the following way
\begin{eqnarray}
R&=& a^0[2(N_{a^{\mu}}+N_{a^0})-s+2] + (a^2 - (a^0)^2)\partial_{a^0} \label{4.13}\\
N_{a^{\mu}}&=& a^{\mu}\partial_{a^{\mu}} ,\label{4.14}\\
N_{a^{0}}&=&a^{0}\partial_{a^{0}} ,\label{4.15}
\end{eqnarray}
with the following important algebraic relations:
\begin{eqnarray}
&&[(N_{a^{\mu}}+N_{a^0}),R]=R ,\label{4.16}\\
&&[(N_{a^{\mu}}+N_{a^{0}}),(a,\hat{\nabla})]=(a,\hat{\nabla}) ,\label{4.17}\\
&&[R,(a,\hat{\nabla})]=2a^{0}(a,\hat{\nabla}) .\label{4.18}
\end{eqnarray}

We should evaluate (\ref{4.11}) on the ground state (\ref{4.2})  $\mid 0>=e^{2(s-1)u}h^{(s)}(x^{\mu};b^{\mu})$, where
\begin{eqnarray}
  a^{\mu}\partial_{a^{\mu}}\mid 0>&=&a^0\partial_{a^0}\mid 0>=0 ,\label{4.19}\\
  R\mid 0> &=&(2-s)a^{0}\mid 0> .\label{4.20}
\end{eqnarray}
Expanding therefore this operator power (\ref{4.11}) into a noncommutative binomial series we get
\begin{eqnarray}
&&[(a,\hat{\nabla})-R]^{n}\mid 0> = \sum_{p=0}^{n}(-1)^{p}\nonumber\\
&&\sum_{n-p \geq i_{p}\geq i_{p-1}\geq i_{p-2}...\geq i_{1} \geq 1}  (a,\hat{\nabla})^{n-p-i_{p}}R(a,\hat{\nabla})^{i_{p}-i_{p-1}}R (a,\hat{\nabla})^{i_{p-1} -i_{p-2}} \dots R(a,\hat{\nabla})^{i_1}\mid 0> .\nonumber\\\label{4.21}
\end{eqnarray}
Then using relation
\begin{equation}\label{4.22}
    [R,(a,\hat{\nabla})^{i}]=2i a^{0}(a,\hat{\nabla}) ,
\end{equation}
we can rewrite (\ref{4.21}) in the following form
\begin{eqnarray}
&&[(a, \hat{\nabla}) - R]^{n} \mid 0> = \sum_{p=0}^{n} (-1)^{p}(a, \hat{\nabla})^{n-p} \nonumber\\
&&\sum_{n-p\geq i_{p}\geq i_{p-1}\geq i_{p-2}...\geq i_1\geq 0}
(2i_{p} a^0 +R)(2i_{p-1}a^0 +R)...  (2i_2a^0+R)(2i_1a^0 +R) \mid 0> .\nonumber\\\label{4.23}
\end{eqnarray}
Now introducing the new notation $\phi_{i_{k}}=2i_{k}a^{0}+R$
and using (\ref{4.13}) and (\ref{4.16}) we obtain
\begin{eqnarray}
&&[(a, \hat{\nabla}) - R]^{n} \mid 0> = \sum_{p=0}^{n} (-1)^{p}(a, \hat{\nabla})^{n-p} \qquad\qquad\qquad \nonumber\\
&&\sum_{n-p\geq i_{p}\geq i_{p-1}\geq i_{p-2}...\geq i_1\geq 0}\phi_{i_{p}}\phi_{i_{p-1}}\dots\phi_{i_{2}}\phi_{i_{1}}\mid 0> ,
\label{4.24}
\end{eqnarray}
where after taking into account that
\begin{equation}\label{4.25}
    [N_{a^{\mu}}+N_{a^0}]\phi_{i_{m}}\dots\phi_{i_{2}}\phi_{i_{1}}\mid 0>
    =m \phi_{i_{m}}\dots\phi_{i_{2}}\phi_{i_{1}}\mid 0> ,
\end{equation}
we have $\phi_{i_{k}}$ as a very simple "creation" operators
\begin{equation}\label{4.26}
    \phi_{i_{k}}=a^{0}[2(i_{k}+k)-s]+[a^{2}-(a^{0})^{2}]\partial_{a^{0}} .
\end{equation}

The summation over the labels $\{i_{k}\}, k \leq p-1$  yields a polynomial in $a^0$ and $(a)^2$ of the form
\begin{equation}
\sum_{k=0}^{\frac{p}{2}} \xi_{k}(i_{p}) (a^2)^{k} (a_0) ^{p-2k} ,
\label{4.27}
\end{equation}
and at the end we have to insert $i_{p} = n-p$ .

\subsection*{ B. The algebra generated by $(a, \nabla^{AdS})$ and $b_0(a,\partial_{b}), (a,b) \partial_{b^0}$}
\setcounter{equation}{0}
\renewcommand{\theequation}{B.\arabic{equation}}
\quad
It remains to evaluate the factors
\begin{eqnarray}
&&(a, \hat{\nabla})^{p} = [(a,\nabla^{AdS}) - (L^{+} +L^{-})]^{p} \nonumber\\
&&= \sum_{n=0}^{p} (-1)^{n}{p \choose n}(a,\nabla^{AdS})^{p-n}(L^{+} + L^{-})^{n} ,
\label{4.28}
\end{eqnarray}
where $L^{+}, L^{-}$ generate a Lie algebra
\begin{eqnarray}
&& L^{+} = b^0(a,\partial_{b}), \quad L^{-} = (a,b)\partial_{b^0} ,\\
&& [L^{+}, L^{-}] = H  = a^2 b^0 \partial_{b^0} -(a,b) (a,\partial_{b}) ,\\
&& [H, L^{\pm}] = \pm 2a^2 L^{\pm} .\qquad\qquad \label{4.31}
\end{eqnarray}
Representations of this Lie algebra are created from an ($s+1$)-dimensional vector space of "null vectors" $\Phi_{n}(a;b)$
of "level" $n$
\begin{equation}
\Phi_{n}(a;b) = h^{(s)}_{\mu_1,\mu_2,...\mu_{s}} a^{\mu_1}a^{\mu_2}... a^{\mu_{n}}b^{\mu_{n+1}}b^{\mu_{n+2}}...b^{\mu_{s}},
\qquad L^{-} \Phi_{n}(a;b) = 0 ,
\label{4.32}
\end{equation}
for any fixed tensor function $h^{s}$. From (\ref{4.31})
follows that starting from $\Phi_{0}(a;b)$ all $\Phi_{n}(a;b)$ can be produced by application of $H$
\begin{eqnarray}
&& H \Phi_0(a;b) = -s (a,b) \Phi_1(a,b) ,\\
&& H^2 \Phi_0(a;b) = [s]_2 (a,b)^2 \Phi_2(a;b)  + s a^2 (a,b) \Phi_1(a;b) ,\\
&& H^3 \Phi_0(a;b) = -\{[s]_3 (a,b)^3 \Phi_3(a;b) + 3[s]_2  a^2 (a,b)^2 \Phi_2(a;b) + s (a^2)^2 (a,b)\Phi_1(a;b)\} .\qquad\qquad
\end{eqnarray}
The ansatz
\begin{equation}
H^{n} \Phi_0(a;b) = (-1)^{n} \{[s]_{n} (a,b)^{n} \Phi_{n}(a;b) +\sum_{r=1}^{n-1} A_{n-r}^{(n)}[s]_{n-r}(a^2)^{r} (a,b)^{n-r}\Phi_{n-r}(a;b)\}, \label{4.36}
\end{equation}
implies the recursion
\begin{eqnarray}
&&A_{r-1}^{(n)} + r A_{r}^{(n)} = A_{r}^{(n+1)} ,\label{4.37}\\
&&A_{r}^{(n)} = 0 \quad\textnormal{for}\quad r>n .
\label{4.38}
\end{eqnarray}
The first order difference equation (\ref{4.37}) has the unique solution
\begin{equation}
A_{n}^{(n+p+1)} = \sum_{k=1}^{n}k A_{k}^{(k+p)} .
\label{4.39}
\end{equation}
The easiest way to exploit eqs. (\ref{4.37})-(\ref{4.39}) is by an ansatz
\begin{equation}
A_{n}^{(n+p)} = \sum_{m=1}^{p} \sigma_{m}^{(p)} {n+p \choose m+p},
\label{4.40}
\end{equation}
where the coefficients $\sigma_{m}^{(p)}$ are natural numbers. These coefficients satisfy recursion relations that can be used to determine them.

With the help of the basis $\{\Phi_{n}(a;b)\}_{n=0}^{s}$ of null vectors the representation of the Lie algebra can be constructed as follows.
We start from
\begin{eqnarray}
&&(L^{+} + L^{-})^{n} \Phi_0(b) = \sum_{p=0}^{n}\sum_{n-p \geq i_{p}\geq i_{p-1}\geq i_{p-2}...\geq i_{1} \geq 1}  \nonumber\\ &&(L^{+})^{n-p-i_{p}} L^{-}(L^{+})^{i_{p}-i_{p-1}}L^{-} (L^{+})^{i_{p-1} -i_{p-2}} L^{-}...(L^{+})^{i_1}\Phi_0(b)
.\label{4.41}
\end{eqnarray}
Only commutators of $L^{-}$ with powers of $L^{+}$ arise
\begin{eqnarray}
&&[L^{-}, (L^{+})^{i}] = -\sum_{k = 0}^{i-1} (L^{+})^{i-k-1}H (L^{+})^k = \qquad \nonumber\\
&&-\sum_{k=0}^{i-1}(L^{+})^{i-1}( H +2k a^2)
=  - (L^{+})^{i-1}( i H + [i]_2\, a^2).
\label{4.42}
\end{eqnarray}
Here we recognize that the whole basis $\{\Phi_{n}(a;b)\}$ of null vectors is produced from $\Phi_0(b)$ by the action of $H$. With the shorthand
\begin{equation}
\psi_{i} = i H + [i]_2\, a^2 ,
\label{4.43}
\end{equation}
the result is
\begin{equation}
\sum_{p=1}^{n}(-1)^{p}(L^{+})^{n-p}\sum_{p=i_{p}\geq i_{p-1}\geq i_{p-2}...\geq i_2 \geq i_1\geq 1} \psi_{i_{p}-p+1}\psi_{i_{p-1}-p+2}
\psi_{i_{p-2}-p+3}...\psi_{i_2-1} \psi_{i_1} \Phi_0(b).
\label{4.44}
\end{equation}
The sum is a homogeneous polynomial of $H$ and $a^2$ of degree $p$, remember that $H$ is second order in $a$ as well.


\begin{thebibliography}{100}
\bibitem{VasilievEqn}
  M.~A.~Vasiliev, ``Consistent equation for interacting gauge fields of all spins in (3+1)-dimensions.'',
  Phys. Lett. B {\bf 243} (1990) 378-382. M.~A.~Vasiliev, `` Nonlinear equations for symmetric massless higher spin fields in $(A)dS_{d}$.''
  Phys. Lett. B {\bf 567} (2003) 139-151, arXiv:hep-th/0304049.
\bibitem{Vasiliev:2012vf}
  M.~A.~Vasiliev,
  arXiv:1203.5554 [hep-th].
\bibitem{Vasiliev:2011zza}
  M.~A.~Vasiliev,
  Phys.\ Usp.\  {\bf 54}, 641 (2011)
  [Usp.\ Fiz.\ Nauk {\bf 181}, 665 (2011)].
\bibitem{Sagnotti:2011qp}
  A.~Sagnotti,
  ``Notes on Strings and Higher Spins,''
  arXiv:1112.4285 [hep-th].
\bibitem{Metsaev}
  R.~R.~Metsaev,
  ``Cubic interaction vertices for massive and massless higher spin fields,''
  Nucl.\ Phys.\  B {\bf 759} (2006) 147
  [arXiv:hep-th/0512342]; R.~R.~Metsaev,  ``Cubic interaction vertices for fermionic and bosonic arbitrary spin  fields,'' arXiv:0712.3526 [hep-th].
\bibitem{MMR1}
  R.~Manvelyan, K.~Mkrtchyan and W.~R\"uhl,
    "General trilinear interaction for arbitrary even higher spin gauge fields",
      Nucl.\ Phys.\  B {\bf 836} (2010) 204, arXiv:1003.2877 [hep-th].
\bibitem{Sagnotti:2010at}
  A.~Sagnotti and M.~Taronna,
  ``String Lessons for Higher-Spin Interactions,''
  Nucl.\ Phys.\ B {\bf 842} (2011) 299
  [arXiv:1006.5242 [hep-th]].
\bibitem{Fotopoulos:2010ay}
  A.~Fotopoulos and M.~Tsulaia,
  JHEP {\bf 1011} (2010) 086
  [arXiv:1009.0727 [hep-th]].
\bibitem{Manvelyan:2010je}
  R.~Manvelyan, K.~Mkrtchyan and W.~R\"uhl,
  Phys.\ Lett.\ B {\bf 696} (2011) 410
  [arXiv:1009.1054 [hep-th]].
\bibitem{Vasilev:2011xf}
  M.~A.~Vasiliev,
  ``Cubic Vertices for Symmetric Higher-Spin Gauge Fields in $(A)dS_d$,''
  arXiv:1108.5921 [hep-th].
\bibitem{polyakov}
Dimitri Polyakov, ``Gravitational Couplings of Higher Spins from String Theory.''
arXiv:1005.5512 [hep-th],
 ``Interactions of Massless Higher Spin Fields From String Theory.''
arXiv:0910.5338 [hep-th].
\bibitem{Bengtsson:1983pd}
  A.~K.~H.~Bengtsson, I.~Bengtsson and L.~Brink,
  Nucl.\ Phys.\ B {\bf 227} (1983) 31.
\bibitem{Bengtsson:1983pg}
  A.~K.~H.~Bengtsson, I.~Bengtsson and L.~Brink,
  Nucl.\ Phys.\ B {\bf 227} (1983) 41.
\bibitem{vanDam1}
  F.~A.~Berends, G.~J.~H.~Burgers and H.~Van Dam,
 ``On Spin Three Selfinteractions,''
 Z.\ Phys.\  C {\bf 24} (1984) 247;
  F.~A.~Berends, G.~J.~H.~Burgers and H.~van Dam,
  ``On The Theoretical Problems In Constructing Interactions Involving Higher
  Spin Massless Particles,''
  Nucl.\ Phys.\  B {\bf 260} (1985) 295.
\bibitem{vanDam}
  F.~A.~Berends, G.~J.~H.~Burgers and H.~van Dam,
  ``Explicit Construction Of Conserved Currents For Massless Fields Of
  Arbitrary Spin,'' Nucl.\ Phys.\  B {\bf 271} (1986) 429;
\bibitem{ouvry}
  I.~G.~Koh, S.~Ouvry, ``Interacting gauge fields of any spin and symmetry,''
  Phys. Lett. B {\bf 179} (1986) 115; Erratum-ibid. {\bf 183} B (1987) 434.
\bibitem{Damour}
 T.~Damour and S.~Deser,
  ``Higher derivative interactions of higher spin gauge fields,''
  Class.\ Quant.\ Grav.\  {\bf 4}, L95 (1987).
\bibitem{Vasiliev}
  E.~S.~Fradkin and M.~A.~Vasiliev, ``On The Gravitational Interaction
  Of Massless Higher Spin Fields,'' Phys.\ Lett.\ B {\bf 189} (1987)
  89.
\bibitem{Vasiliev1}
  E.~S.~Fradkin and M.~A.~Vasiliev, ``Cubic Interaction In
  Extended Theories Of Massless Higher Spin Fields,'' Nucl.\ Phys.\ B
  {\bf 291} (1987) 141.
\bibitem{Metsaev:1993ap}
  R.~R.~Metsaev,
  ``Generating function for cubic interaction vertices of higher spin fields in any dimension,''
  Mod.\ Phys.\ Lett.\ A {\bf 8} (1993) 2413.
\bibitem{Vasiliev2}
  M.~A.~Vasiliev, "Cubic Interactions of Bosonic Higher Spin Gauge Fields in $AdS_{5}$",
  [arXiv:hep-th/0106200].
  K.~B.~Alkalaev,  M.~A.~Vasiliev, "N = 1 Supersymmetric Theory of Higher Spin Gauge Fields in $AdS_{5}$ at the Cubic Level",
  [arXiv:hep-th/0206068]
\bibitem{Manvelyan:2004mb}
  R.~Manvelyan and W.~R\"uhl,
  ``Conformal coupling of higher spin gauge fields to a scalar field in AdS(4) and generalized Weyl invariance,''
  Phys.\ Lett.\ B {\bf 593} (2004) 253
  [hep-th/0403241].
\bibitem{boulanger}
  Nicolas Boulanger, Serge Leclercq, Per Sundell,
  ``On The Uniqueness of Minimal Coupling in Higher-Spin Gauge Theory,''
  JHEP 0808:056,2008;  [arXiv:0805.2764 [hep-th]].
\bibitem{bek}
Xavier Bekaert, Nicolas Boulanger and Serge Leclercq,
``Strong obstruction of the Berends-Burgers-van Dam spin-3 vertex.''
J.Phys.A43:185401,2010. arXiv:1002.0289 [hep-th].
Xavier Bekaert, Nicolas Boulanger, Sandrine Cnockaert, Serge Leclercq,
  ``On Killing tensors and cubic vertices in higher-spin gauge theories,''
  Fortsch. Phys. {\bf 54} (2006) 282-290; [arXiv:hep-th/0602092].
\bibitem{Petkou}
  A.~Fotopoulos, N.~Irges, A.~C.~Petkou and M.~Tsulaia,
  ``Higher-Spin Gauge Fields Interacting with Scalars: The Lagrangian Cubic
  Vertex,''
  JHEP {\bf 0710} (2007) 021;
  [arXiv:0708.1399 [hep-th]].
  I.~L.~Buchbinder, A.~Fotopoulos, A.~C.~Petkou and M.~Tsulaia,
  ``Constructing the cubic interaction vertex of higher spin gauge fields,''
  Phys.\ Rev.\  D {\bf 74} (2006) 105018;
  [arXiv:hep-th/0609082].
    A. Fotopoulos and M. Tsulaia,
    ``Current Exchanges for Reducible Higher Spin Modes on AdS.'' arXiv:1007.0747 [hep-th]
\bibitem{Manvelyan:2009tf}
  R.~Manvelyan and K.~Mkrtchyan,
  ``Conformal invariant interaction of a scalar field with the higher spin field in AdS(D),''
  Mod.\ Phys.\ Lett.\ A {\bf 25} (2010) 1333
  [arXiv:0903.0058 [hep-th]].
\bibitem{M}
  R.~Manvelyan, K.~Mkrtchyan and W.~R\"uhl,
  ``Off-shell construction of some trilinear higher spin gauge field
  interactions,''
  Nucl.\ Phys.\  B {\bf 826} (2010) 1
  [arXiv:0903.0243 [hep-th]].
\bibitem{MMR2}
  R.~Manvelyan, K.~Mkrtchyan and W.~R\"uhl,
  ``Direct construction of a cubic selfinteraction for higher spin gauge fields,''
  Nucl.\ Phys.\ B\ {\bf 844} (2011) 348
  [arXiv:1002.1358 [hep-th]].
\bibitem{Taronna:2010qq}
  M.~Taronna,
  ``Higher Spins and String Interactions,''
  arXiv:1005.3061 [hep-th].
\bibitem{Zinoviev}
Yu.M. Zinoviev, ``Spin 3 cubic vertices in a frame-like formalism.''
JHEP 1008:084,2010; arXiv:1007.0158 [hep-th]
\bibitem{Mkrtchyan:2010pp}
  K.~Mkrtchyan,
  ``Higher Spin Interacting Quantum Field Theory and Higher Order Conformal Invariant Lagrangians,''
  arXiv:1011.0160 [hep-th].
\bibitem{Mkrtchyan:2010zz}
  K.~Mkrtchyan,
  ``Linearized interactions of scalar and vector fields with the higher spin field in AdSD,''
  Armenian J.\ Phys.\  {\bf 3} (2010) 98
   [Phys.\ Part.\ Nucl.\ Lett.\  {\bf 8} (2011) 266].
\bibitem{Mkrtchyan:2011uh}
  K.~Mkrtchyan,
  ``On generating functions of Higher Spin cubic interactions,''
  arXiv:1101.5643 [hep-th].
\bibitem{Taronna:2011kt}
  M.~Taronna,
  JHEP {\bf 1204} (2012) 029
  [arXiv:1107.5843 [hep-th]].
\bibitem{Ruehl:2011tk}
  W.~R\"uhl,
  ``Solving Noether's equations for gauge invariant local Lagrangians of N arbitrary higher even spin fields,''
  arXiv:1108.0225 [hep-th].
\bibitem{Joung:2011ww}
  E.~Joung and M.~Taronna,
  ``Cubic interactions of massless higher spins in (A)dS: metric-like approach,''
  Nucl.\ Phys.\ B {\bf 861} (2012) 145
  [arXiv:1110.5918 [hep-th]].
\bibitem{Tsulaia:2012rb}
  M.~Tsulaia,
  ``On Tensorial Spaces and BCFW Recursion Relations for Higher Spin Fields,''
  Int.\ J.\ Mod.\ Phys.\ A {\bf 27}, 1230011 (2012)
  [arXiv:1202.6309 [hep-th]].
\bibitem{Joung:2012rv}
  E.~Joung, L.~Lopez and M.~Taronna,
  ``On the cubic interactions of massive and partially-massless higher spins in (A)dS,''
  arXiv:1203.6578 [hep-th].
\bibitem{Dempster:2012vw}
  P.~Dempster and M.~Tsulaia,
  ``On the Structure of Quartic Vertices for Massless Higher Spin Fields on Minkowski Background,''
  arXiv:1203.5597 [hep-th].
\bibitem{Metsaev:2012uy}
  R.~R.~Metsaev,
  ``BRST-BV approach to cubic interaction vertices for massive and massless higher-spin fields,''
  arXiv:1205.3131 [hep-th].
\bibitem{Bekaert:2010hk}
  X.~Bekaert and E.~Meunier,
  ``Higher spin interactions with scalar matter on constant curvature spacetimes: conserved current and cubic coupling generating functions,''
  JHEP {\bf 1011} (2010) 116
  [arXiv:1007.4384 [hep-th]].
\bibitem{Bekaert:2003uc}
  X.~Bekaert, I.~L.~Buchbinder, A.~Pashnev and M.~Tsulaia,
  ``On higher spin theory: Strings, BRST, dimensional reductions,''
  Class.\ Quant.\ Grav.\  {\bf 21} (2004) S1457
  [hep-th/0312252].
\bibitem{Francia:2008hd}
  D.~Francia, J.~Mourad and A.~Sagnotti,
  ``(A)dS exchanges and partially-massless higher spins,''
  Nucl.\ Phys.\ B {\bf 804} (2008) 383
  [arXiv:0803.3832 [hep-th]].
\bibitem{Alkalaev:2011zv}
  K.~Alkalaev and M.~Grigoriev,
  ``Unified BRST approach to (partially) massless and massive AdS fields of arbitrary symmetry type,''
  Nucl.\ Phys.\ B {\bf 853} (2011) 663
  [arXiv:1105.6111 [hep-th]].
\bibitem{Alkalaev:2009vm}
  K.~B.~Alkalaev and M.~Grigoriev,
  ``Unified BRST description of AdS gauge fields,''
  Nucl.\ Phys.\ B {\bf 835} (2010) 197
  [arXiv:0910.2690 [hep-th]].
\bibitem{Kleb}
  I.~R.~Klebanov and A.~M.~Polyakov, ``AdS dual of the critical O(N)
  vector model,'' Phys.\ Lett.\ B {\bf 550} (2002) 213;
  [arXiv:hep-th/0210114].
\bibitem{Giombi:2011ya}
  S.~Giombi and X.~Yin,
  arXiv:1105.4011 [hep-th].
\bibitem{Maldacena:2012sf}
  J.~Maldacena and A.~Zhiboedov,
  arXiv:1204.3882 [hep-th].
\bibitem{Campoleoni:2011hg}
  A.~Campoleoni, S.~Fredenhagen and S.~Pfenninger,
  ``Asymptotic W-symmetries in three-dimensional higher-spin gauge theories,''
  JHEP {\bf 1109} (2011) 113
  [arXiv:1107.0290 [hep-th]].
\bibitem{Banados:2012ue}
  M. Banados, R.~Canto and S.~Theisen,
  ``The action for higher spin black holes in three dimensions,''
  arXiv:1204.5105 [hep-th].
\bibitem{Henneaux:2010xg}
  M.~Henneaux and S.~-J.~Rey,
  ``Nonlinear $W_{\infty}$ as Asymptotic Symmetry of Three-Dimensional Higher Spin Anti-de Sitter Gravity,''
  JHEP {\bf 1012} (2010) 007
  [arXiv:1008.4579 [hep-th]].
\bibitem{Gutperle:2011kf}
  M.~Gutperle and P.~Kraus,
  ``Higher Spin Black Holes,''
  JHEP {\bf 1105} (2011) 022
  [arXiv:1103.4304 [hep-th]].
\bibitem{Kraus:2011ds}
  P.~Kraus and E.~Perlmutter,
  ``Partition functions of higher spin black holes and their CFT duals,''
  JHEP {\bf 1111} (2011) 061
  [arXiv:1108.2567 [hep-th]].
\bibitem{BS}
  T.~Biswas and W.~Siegel,
  ``Radial dimensional reduction: Anti-de Sitter theories from flat,''  JHEP\ {\bf 0207} (2002) 005  [hep-th/0203115].
\bibitem{Waldron}
  K.~Hallowell and A.~Waldron,
  ``Constant curvature algebras and higher spin action generating functions,''  Nucl.\ Phys.\ B\ {\bf 724} (2005) 453  [hep-th/0505255].
\bibitem{Frons}
    C.~Fronsdal, ``Singletons And Massless, Integral Spin Fields On De Sitter Space (Elementary
    Particles In A Curved Space Vii),'' Phys.\ Rev.\ D {\bf 20},
    (1979) 848;``Massless Fields With Integer Spin,'' Phys.\ Rev.\ D {\bf 18} (1978) 3624.
\bibitem{MMR5}
R.~Manvelyan, K.~Mkrtchyan and W.~R\"uhl,
  ``Ultraviolet behaviour of higher spin gauge field propagators and one loop
  mass renormalization,''
  Nucl.\ Phys.\  B {\bf 803} (2008) 405
  [arXiv:0804.1211 [hep-th]].
\end{thebibliography}
\end{document}